\documentclass[apj]{emulateapj}
\slugcomment{Accepted to ApJ. Draft version November 17, 2016}

\usepackage{aas_macros}
\usepackage{listings}
\usepackage{amsmath}
\usepackage{amssymb}
\usepackage{graphicx,graphics}
\usepackage{rotating}
\usepackage{multirow}
\usepackage{natbib}
\usepackage{color}
\usepackage{soul}
\usepackage{apjfonts}
\usepackage{hyperref}
\hypersetup
{
    pdfauthor={Kazuma Mitsuda},
    pdftitle={Isophote Shapes of Early-Type Galaxies in Massive Clusters at \lowercase{$z\sim1$} and \lowercase{$0$}},
}
\shorttitle{Isophote shapes of ETGs in clusters at $z\sim1$ and $0$}
\shortauthors{Mitsuda et al.}

\begin{document}

\title{Isophote Shapes of Early-Type Galaxies in Massive Clusters at \lowercase{$z\sim1$} and \lowercase{$0$}}

\author{Kazuma Mitsuda \altaffilmark{1,2}}
\author{Mamoru Doi \altaffilmark{2, 3, 4}}
\author{Tomoki Morokuma \altaffilmark{2, 3}},
\author{Nao Suzuki \altaffilmark{3}},
\author{Naoki Yasuda \altaffilmark{3}},
\author{Saul Perlmutter \altaffilmark{5,6}},
\author{Greg Aldering \altaffilmark{6}},
\author{Joshua Meyers \altaffilmark{7}}
\affil{${}^{1}$\ Department of Astronomy, School of Science, 
The University of Tokyo, 7-3-1 Hongo, Bunkyo-ku, Tokyo 113-0033, Japan}
\affil{${}^{2}$\ Institute of Astronomy, School of Science, The University of Tokyo,
2-21-1 Osawa, Mitaka, Tokyo 181-0015, Japan;
\href{mailto:kazuma@ioa.s.u-tokyo.ac.jp}{kazuma@ioa.s.u-tokyo.ac.jp}}
\affil{${}^{3}$\ Kavli IPMU (WPI), UTIAS, The University of Tokyo, Kashiwa, Chiba 277-8583, Japan}
\affil{${}^{4}$\ Research Center for the Early Universe, the University of Tokyo,
7-3-1 Hongo, Bunkyo-ku, Tokyo 113-0033, Japan}
\affil{${}^{5}$\ Department of Physics, University of California Berkeley, Berkeley, CA 94720, USA}
\affil{${}^{6}$\ E.O. Lawrence Berkeley National Lab, 1 Cyclotron Rd., Berkeley, CA 94720, USA}
\affil{${}^{7}$\ Department of Physics, Stanford University, 450 Serra Mall, Stanford, CA 94305, USA}

\begin{abstract}

We compare the isophote shape parameter $a_{4}$ of early-type galaxies (ETGs)
between $z\sim1$ and 0 as a proxy for dynamics to investigate the epoch at which the dynamical properties of ETGs are established,
using cluster ETG samples with stellar masses of $\log(M_{*}/M_{\odot})\geq10.5$
which have spectroscopic redshifts.
We have 130 ETGs  from the {\it Hubble Space Telescope} Cluster Supernova Survey for $z\sim1$
and 355 ETGs from the {\it Sloan Digital Sky Survey} for $z\sim0$.
We have developed an isophote shape analysis method which can be used for high-redshift galaxies and has been carefully compared with published results.
We have applied the same method for both the $z\sim1$ and $0$ samples.
We find similar dependence of the $a_{4}$ parameter on the mass and size at $z\sim1$ and 0;
the main population of ETGs changes from disky to boxy at a critical stellar mass of $\log(M_{*}/M_{\odot})\sim11.5$
with the massive end dominated by boxy.
The disky ETG fraction decreases with increasing stellar mass both at $z\sim1$ and $0$,
and 
is consistent between these redshifts in all stellar mass bins when the Eddington bias is taken into account.
Although uncertainties are large,
the results suggest that the isophote shapes and probably dynamical properties of ETGs in massive clusters are already in place at $z>1$
and do not significantly evolve in $z<1$,
despite significant size evolution in the same galaxy population.
The constant disky fraction favors less violent processes than mergers
as a main cause of the size and morphological evolution of intermediate mass ETGs in $z<1$.
\end{abstract}

\setcounter{section}{0}

\section{Introduction}
Understanding the formation and evolution of early-type galaxies (ETGs) is one of the main topics in the modern astronomy
as they are important ingredients in the universe.
ETGs, also referred to as ellipticals and S0s,
are one of the major galaxy populations.
The evolution histories of ETGs are imprinted in dynamics, stellar populations, and shapes.

Dynamics of ETGs provides crucial knowledge about their formation and evolution histories.
There is a well-known parameter correlation between the luminosity, velocity dispersion, and size,
i.e., the Fundamental Plane \citep{Djorgovski+87},
which combines the correlation between the total luminosity and velocity dispersion \citep{FaberJackson76},
and that between the size and surface brightness \citep{Kormendy77}.
Using integral field spectroscopic (IFS) data of 260 local ETGs \citep{Cappellari+11},
\citet{Cappellari+13a} obtain robust stellar mass estimator,
and confirm the mass-to-luminosity ratio ($M/L$) variation as a function of velocity dispersion,
with which the Fundamental Plane can be interpreted as virial equilibrium \citep[e.g.,][]{Djorgovski+87, PrugnielSimien96, Forbes+98},
which implies the Fundamental Plane can be reduced to the mass-size plane \citep[][for a review]{Cappellari15IAU}.

\citet{Emsellem+07, Emsellem+11} analyze the IFS data and classify ETGs, using the specific angular momentum,
into fast rotators whose dynamics is dominated by rotation
and dispersion-dominated slow rotators.
Considered on the mass-size plane, the specific angular momentum of ETGs {\it varies} as a function of constant velocity dispersion assuming virial relation.
The massive end of ETGs is dominated by slow rotators wheres lower mass systems are basically fast rotators \citep[see Figure 8 in][]{Cappellari+13b}.
The similar trend is also found in the tridimensional structure, i.e., massive ETGs are spherical whereas less massive ones are flattened, oblate spheroids \citep[Figure 7 in][]{Cappellari+13b}.

The stellar population is also important to discuss the formation and evolution of ETGs.
Local ETGs are known to populate a tight red sequence in the color-magnitude or color-stellar mass diagram \citep{Baum59, Faber73, Visvanathan_Sandage77, Baldry+04, Baldry+06},
as they have rather homogeneous old and metal-rich stellar populations \citep[e.g.,][]{Bower+92, KodamaArimoto97}.
Analyzing stellar absorption features obtained from galaxy spectra have revealed that
more massive ETGs are older, more metal-rich, and have more $\alpha$ element enhancement
which implies shorter star formation time-scales \citep{Worthey+92, Thomas+05, Thomas+10}.
Considered on the mass-size plane, unlike dynamical properties,
the stellar population parameters such as the stellar age, metallicity, and star formation time-scales \citep{McDermid+15}
as well as molecular gas fraction \citep{Young+11, Cappellari+13b} are {\it constant} as a function of constant velocity dispersion.
On the other hand, with increasing velocity dispersion, 
the stellar populations become older, more metal-rich and more $\alpha$ element enhanced with shorter formation time-scales,
and galaxies have less molecular gas fraction \citep{Cappellari+13b, McDermid+15}.
As bulge fraction also increases with increasing velocity dispersion \citep[][and references therein]{Cappellari+13b}, 
characterization of stellar population properties are probably linked to the bulge formation.

To explain the different correlation of dynamics and stellar population to the velocity dispersion,
two-phase formation scenario is favored, i.e.,
a massive compact fast-rotating bulge is formed by dissipative processes such as gas inflow or wet mergers at high redshift ($z>2$) when the universe is much more gas rich,
and dissipationless processes such as dry minor or major mergers increase the galaxy size
\citep[e.g., ][and references therein]{Khochfar+11, Cappellari+13b, DekelBurkert14}.
As dry mergers reduce angular momentum of the fast-rotating bulge and alter it into slowly-rotationg ETGs \citep{Khochfar+05, Naab+06} without changing the stellar population,
the dynamical properties varies in the mass-size plane as a function of constant velocity dispersion with the massive end of ETGs dominated by slow rotators
while the stellar population becomes constant.

\citet{Khochfar+11} investigate the evolution of the ratio of fast to slow rotators using semi-analytic galaxy formation model,
and present that the ratio evolves in $z<2$ such that the fast to slow ratio decreases with decreasing redshift due to dry mergers.
The size evolution observed in $z<2$ supports the hypothesis that dry (mainly minor) mergers are at work in this redshift range \citep[e.g.,][]{Trujillo+06, Trujillo+07, van_Dokkum+08, Damjanov+09, Barro+13, Tadaki+14}.
However, \citet{Naab+14} present, using cosmological hydrodynamical simulations of individual galaxies, that there are many paths to create fast- and slowly-rotating ETGs and
it is not clear when and how ETGs obtained their dynamical properties after the star formation quenching at $z\gtrsim2$.
Moreover in $z<2$, secular processes such as fading of disks may play an important role,
if we take account of the insufficient merger rate and the evolution of morphology as well as the size evolution
\citep[e.g.,][]{Oesch+10, Carollo+13, Carollo+14aph, De_Propris+15}.

It is important to study the evolution of the dynamical properties of ETGs observationally.
However, it is difficult to carry out IFS observations for high-redshift ETGs even with 8-meter class telescopes.
The shapes of ETGs bring us important information which is related to the dynamics.
Although ETGs look featureless and their isophote shapes can be described by perfect ellipses to the zero-th order,
their shapes have small but significant deviation from ellipses into ``boxy'' or ``disky'' \citep{Lauer85c, Lauer85a, Lauer85b}.
\citet{Bender+87} and \citet{Jedrzejewski87} evaluated the deviations of isophote shapes from perfect ellipses
using Fourier expansions in the polar angle.
They found the most significant non-zero component is the $a_4$ parameter, the coefficient of the $\cos (4\theta)$ term.
The negative sign of the parameter indicates the boxy isophote
whereas the positive sign indicates  disky.
\citet{Bender+88, Bender+89} study the isophote shapes and relation to other observed properties using 69 bright E/S0 sample,
and have shown that there are significant correlations between $a_4$ and radio, X-ray properties.
Boxy ETGs tend to be brighter, supported by random motions with large velocity anisotropy (i.e., slow rotators),
have significant radio and X-ray activities and core nuclear light profiles,
while disky ETGs tend to be fainter,
supported by ordered rotation with small velocity anisotropy (i.e., fast rotators),
lack radio and X-ray activities and have coreless nuclear profiles
\citep{Bender+89, Ferrarese+94, van_den_Bosch+94, Lauer+95, Kormendy+96, Faber+97, Rest+01, Lauer+05, Kormendy+09}.
Dependence of the isophote shapes on environment is also studied by \citet{Shioya+93}.
These isophote shape properties are confirmed by \citet{Hao+06} with much larger sample of 847 local ETGs
using the {\it Sloan Digital Sky Survey} (SDSS) data.
\citet{Pasquali+07} use the same sample to conclude that isophote shapes may be related to nuclear activities and to group hierarchy.
\citet{Naab+99} and \citet{Naab+03} present, by numerical simulations, nearly equal mass mergers between disk galaxies produce boxy ellipticals
whereas minor mergers result in disky ones.
\citet{Khochfar+05} and \citet{Naab+06} also show that dry mergers also produce boxy elliptical galaxies
regardless of the progenitor mass ratio.

In this study, to investigate the epoch at which the dynamical properties of ETGs are established,
we analyze the isophote shape parameter ($a_{4}$), as a proxy for the dynamical properties, of ETGs in massive clusters at $z\sim1$ and 0.
We compare the dependence of the isophote shape parameter on the mass and size as well as the disky ETG fraction between these redshifts.
Advantages of studying galaxies in massive clusters are that there are larger number of ETGs than in fields,
massive clusters are unique environment which harbors massive ETGs such as central and cD galaxies,
and galaxies evolve within the cluster once they enter into such an environment. 
We have created quiescent ETG samples in massive galaxy clusters with spectroscopic redshifts,
using data obtained in the {\it Hubble Space Telescope} ({\it HST}) Cluster Supernova (SN) Survey for $z\sim1$ \citep[][PI-Perlmutter: GO-10496]{Dawson+09},
and SDSS \citep{York+00} Data Release 12 \citep[DR12,][]{Alam+15} for $z\sim0$.
We have also developed an isophote analysis method optimized to high-redshift galaxies with low surface brightness and small apparent size.

This paper is organized as follows:
In Section \ref{Sec: Sample}, we describe the sample selection, and basic properties of the high- and low-redshift quiescent ETG samples.
In Section \ref{Sec: method}, we describe the isophote analysis method.
In Section \ref{Sec: Results}, we present the results which are followed by discussion in Section \ref{Sec: Discuss}.
Throughout this paper, magnitudes are described in the AB system and are galactic extinction corrected \citep{Schlegel+98, Schlafly+11}.
We assume a $\mathrm{\Lambda}$CDM cosmology with parameters of ($\Omega_m$, $\Omega_{\Lambda}$, $H_0$)=(0.3, 0.7, 70 $\mathrm{km\ s^{-1}\ Mpc^{-1}}$).


\section{The Galaxy Samples}
\label{Sec: Sample}
In this section, we describe the sample selection and basic properties of our sample galaxies.
We create a stellar-mass limited, high-redshift quiescent ETG sample and low-redshift counterpart for comparison.
We first select galaxies with spectroscopic redshift from the {\it HST} Cluster SN Survey \citep{Dawson+09} for the high redshift ($z\sim1$),
and from SDSS \citep{York+00} DR12 \citep{Alam+15} for the low redshift ($z\sim0$).
To select quiescent ETGs, we then choose quiescent galaxies using color magnitude diagram,
impose a stellar mass limit, and select ETGs based on morphological parameters.

\subsection{High-Redshift Galaxy Sample}
\label{Subsec: Hiz_sample}
In the {\it HST} Cluster SN Survey \citep{Dawson+09} survey,
twenty-five massive high-redshift clusters have been selected from X-ray, optical, and IR suyveys \citep{Dawson+09}.
The basic properties of the clusters such as redshifts, virial masses, and radiii are described in \citet{Jee+11}.
We have obtained multi-epoch {\it HST} imaging data (PID 10496)
and follow-up spectroscopic data of galaxies in the clusters.

\subsubsection{{\it HST} Imaging Data}
Imaging data obtained by {\it HST} are described in \citet{Suzuki+12} and \citet{Meyers+12},
but we briefly describe basic information here.
The twenty-five target clusters were visited by {\it HST} four to nine times between July 2005 and December 2006. 
Each visit typically consisted of four $\sim$500 s exposures in the F850LP filter (hereafter $z_{850}$)
of the Advanced Camera for Surveys (ACS) Wide Field Camera (WFC) \citep{ACS_book}
and one $\sim$500 s exposure in the F775W filter (hereafter $i_{775}$) of the ACS WFC.
For galaxies at $z\sim1$, 
these two photometric bands cover the wavelength region around the 4000\AA\ break, an important spectral feature of quiescent galaxies.

In this paper we use the deep co-additions of exposures from all observation epochs.
Four clusters, RDCS J0910+54 \citep{Mei+06a}, RDCS J0848+44 \citep{Postman+05}, RDCS J1252-29 \citep{Blakeslee+03a},
and XMMU 2235.3-2557 \citep{Jee+09}, had been previously targeted by ACS in $i_{775}$ and $z_{850}$ (PID9290 and PID9919),
and these exposures are also included in our co-added images.
We create cutouts of $i_{775}$ and $z_{850}$ images of each galaxy from the co-additions.
We use $z_{850}$ cutouts for the morphological classification as well as the isophote shape analysis
since the co-added images in $z_{850}$ is much deeper (effective exposure time is $\sim$10k sec or more depending on clusters) than in $i_{775}$.

\subsubsection{Spectroscopic Redshifts}
We select the cluster members confirmed by spectroscopic redshifts.
The redshifts of the galaxies are taken from a spectroscopic catalog created in the {\it HST} Cluster SN Survey \citep{Meyers+12}.
The catalog information is described in \citet{Meyers+12}.
Briefly, as the {\it HST} Cluster SN Survey produced SN candidates, 
galaxies were spectroscopically targeted with multi-object slits
using prescheduled observing time on DEIMOS on Keck II \citep{Faber+03},
and FOCAS on Subaru \citep{Kashikawa+02},
and with Target of Opportunity (ToO) requests on FORS1 and FORS2 on Kueyen and Antu at the Very Large Telescope \citep{Appenzeller+98}. 
The FORS1, FORS2, and DEIMOS observations are described in \citet{Lidman+05} and \citet{Dawson+09};
the FOCAS observations are described in \citet{Morokuma+10}.
Galaxy redshifts are measured through cross-correlation with template eigenspectra derived from SDSS spectra \citep{Aihara+11}.
The important spectroscopic features are the 4000\AA\ break, the absorption of Ca H, K,
and the emission lines of [\ion{O}{2}] 3727\AA\ doublet.
The spectroscopic catalog includes these redshifts and additional ones from literature \citep{Andreon+08b, Bremer+06, Brodwin+06, Demarco+07, Eisenhardt+08, Hilton+07, Hilton+09, Postman+98, Rosati+99, Stanford+02, Stanford+05}.
The equivalent width (EW) of the [\ion{O}{2}] is also provided for some galaxies (about a half of $z\sim1$ cluster galaxies).
Note that the completeness of the spectroscopic sample is not high and varies with cluster to cluster,
since many galaxies are additional targets in {\it HST} Cluster SN Survey whose main targets are SNe and their hosts.

In this study, we include nineteen clusters with more than two spectroscopically identified members,
and exclude other clusters due to too few spectroscopic members.
The total mass of the included clusters spans from $\log(M_{200}/M_{\odot})\sim14.2$ to $14.9$,
where $M_{200}$, adopted from \citet{Jee+11}, is the total mass at the radius, $R_{200}$, inside of which the mean density is 200 times the critical density of the universe at the cluster redshift.
We have 301 $z\sim1$ cluster galaxies with the spectroscopic redshift in total at this stage.
The redshift of the selected galaxies spans from $0.90$ to $1.48$ with the median redshift of $z \sim 1.2$.
We later create quiescent ETGs from these 301 galaxies.
Of these galaxies, 286 lie within one $R_{200}$ from the cluster center, and other 15 galaxies within 1.5 $R_{200}$,
where $R_{200}$ is adopted from \citet{Jee+11}.
The redshifts of 279 galaxies are within $\pm0.01$ from the cluster redshift, and those of other 22 are within $\pm0.02$.

\subsection{Low-Redshift Galaxy Sample}
\label{Subsec: Loz_sample}
To create the low-redshift sample, we make use of SDSS public DR12 \citep{Alam+15}.
We refer to the spectroscopic and imaging catalogs provided by SDSS to select low-redshift galaxies,
and we use $g$-band images for the morphological classification as well as the isophote shape analysis
as it covers the similar rest-frame wavelength range to $z_{850}$ for $z\sim1$ galaxies.
We create a cutout of $g$-band image of each galaxy.

\subsubsection{Low-Redshift Massive Clusters}
We selected nine low-redshift massive clusters which may be possible descendants of the high-redshift ones based on halo masses ($M_{200}$) and redshifts.
%
\citet{Reiprich+02} study basic properties of low-redshift galaxy clusters such as mass and radius based on X-ray observations.
We first select ten clusters whose redshifts lie in the range $0.02< z_{\mathrm{CL}} <0.05$.
The redshift range is determined so that the PSF size of SDSS images for $z\sim0$ galaxies become comparable to
that of {\it HST} ACS $z_{850}$ images for $z\sim1$ galaxies in physical scales in order to match the effect of PSF on isophote shape measurements (see Subsection \ref{subsec: eff_im_qual}).
The PSF FWHM of {\it HST} ACS $z_{850}$ images, $\sim0\farcs1$, corresponds to $\sim0.75-0.85$ kpc at $z\sim0.8-1.5$
whereas that of SDSS images, $\sim1\farcs3$, corresponds to $\sim0.5-1.3$ kpc at $z\sim0.02-0.05$.

Then, one low-mass cluster, MKW4 ($\log(M_{200} / M_{\odot})$=14.1) is excluded from the low-redshift cluster sample.
Since the high-redshift clusters are massive ($\log(M_{200} / M_{\odot})\sim14.2-14.9$), such a low-mass cluster is not likely to be a descendant of the high-redshift ones.
The redshift and mass selection leave us nine clusters, 
A0119, A1367, COMA, MKW8, A2052, MKW3S, A2063, A2147, A2199.
The halo mass spans from $\log(M_{200}/M_{\odot}) \sim 14.6$ to $15.2$.
The masses of low-redshift clusters are slightly larger than those of the high-redshift ones.
Since the high-redshift clusters with the halo mass of $\sim10^{14.5}M_{\odot}$ at $z\sim1$
will evolve into clusters with the mass of $\sim10^{15}M_{\odot}$ at $z\sim0$,
considering halo mass growth from z$\sim$1 to 0 \citep[e.g., ][]{Zhao+09},
the low-redshift clusters are the possible descendants of the high-redshift ones.

\subsubsection{Selection of the Low-Redshift Galaxies}
We select member galaxies of each cluster using SDSS spectroscopic catalog.
From the spectroscopic catalog, we select all galaxies that lie within one $R_{200}$ radius from the cluster center
and within a redshift range of $z_{\mathrm{CL}} - \Delta z \leq z \leq z_{\mathrm{CL}} + \Delta z$,
where we set $\Delta z = 0.0067$ which corresponds to $2000\ \mathrm{km \cdot s^{-1}}$.
The cluster radius ($R_{200}$), center, and redshift are referred from \citet{Reiprich+02}.
We have 3278 galaxies with SDSS spectroscopy ($r < 17.77$ mag) in total at this point with median redshift of $z\sim0.029$.

\subsection{Selection of the Quiescent Galaxies}
\label{Subsec: Select QGs}
We select quiescent galaxies from the spectroscopic members of the high- and low-redshift clusters.
We simply select red galaxies for the high-redshift sample.
For the low-redshift, we impose more strict selection criteria based on the color magnitude diagram
to choose red-sequence galaxies which are possible descendants of the high-redshift red galaxies.

\subsubsection{High-Redshift Quiescent Galaxies}
\label{Subsubsec: Hiz QGs}
We select the high-redshift quiescent galaxies based on their $i_{775}-z_{850}$ colors.
The $i_{775}-z_{850}$ color is measured within a circular aperture of a fixed size.
As galaxies, both late- and early-type galaxies \citep[e.g.,][]{Tortora+10, denBrok+11}, often have radial color gradients, we measure the color only in a central region.
We set the aperture diameter to $0\farcs22$ which corresponds to 1.8 kpc at $z=1.2$ in physical scale.
This physical scale is comparable to that of the diameter of the SDSS 3-arcsec fiber at $z\sim0.03$
with which we later measure the color of low-redshift galaxies.

As the size of PSF is slightly different between $i_{775}$ and $z_{850}$ images,
we prepare PSF-matched images by convolving $i_{775}$ with the PSF of $z_{850}$ and vice versa.
The PSF of each band is created in each cluster field as an averaged image of stars.
We select $\sim30$ unsaturated stars with , cut out 100$\times$100 pixels around them, and normalize the flux with the central value.
Then we oversample the cut-out images by $51\time51$ times, aligned the center in the subpixel level, and take an average.

\begin{figure*}[htbp]
	\vspace{0\baselineskip}
	\centering
	\hspace{-2.0cm}	
	\includegraphics[width=11.0cm]{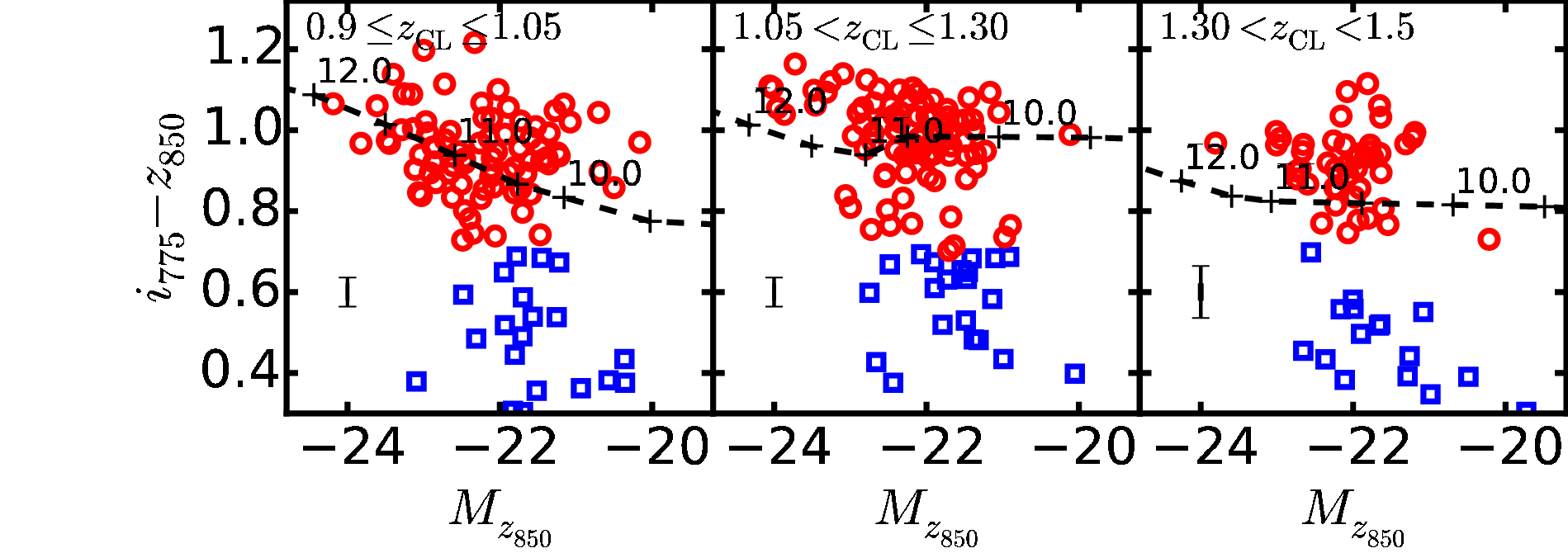}
	\caption{
		Color-magnitude diagrams of the high-redshift galaxies in three cluster redshift bins, 
		$0.9 \leq z_{CL}\leq1.05$, $1.05<z_{CL}\leq1.30$, and $1.30<z_{CL} < 1.5$, from left to right.
		{\it Red circles} represent selected galaxies as quiescent while {\it blue squares} are unselected galaxies.
		Error bars indicate median uncertainties.
		{\it Dashed lines} represent color-magnitude relation of the SSP models with the numbers at the ticks standing for the stellar mass, $\log(M_{*} / M_{\odot})$.
		\label{fig: CMD_Hiz}
	}
	\vspace{0\baselineskip}
\end{figure*}

We select red galaxies with $i_{775}-z_{850} \geq 0.7$ as quiescent galaxies.
The number of galaxies in the high-redshift quiescent galaxy sample is 224.
In FIgure \ref{fig: CMD_Hiz}, we show the color-magnitude diagrams of the high-redshift galaxies in three cluster redshift bins, $z_{CL}\leq1.05$, $1.05<z_{CL}\leq1.30$, and $1.30<z_{CL}$.
We derive the measurement errors of $i_{775}-z_{850}$ color from unconvolved $i_{775}$ and $z_{850}$ images to avoid correlated noise.
We run \verb"SExtractor" \citep{Bertin+96} on cutout images of $i_{775}$ and $z_{850}$ using two-step (Cold/Hot) method \citep{Rix+04} to derive total $z_{850}$ magnitudes.
The detection parameters in the two steps are optimized by trial and error judged by the successful identification and segmentation of galaxies near cluster cores.
We use the Petrosian \citep{Petrosian+76} magnitudes (\verb"MAG_PETRO" in \verb"SExtractor") as the total $z_{850}$ magnitude
as we refer to the the Petrosian magnitudes given in the SDSS catalog for the low-redshift galaxies.

In FIgure \ref{fig: CMD_Hiz}, we also plot the color-magnitude relation of simple stellar population (SSP) models using \citet[hereafter BC03]{Bruzual+03} with Salpeter initial mass function (IMF).
We compute absolute $z_{850}$ magnitudes and $i_{775}-z_{850}$ colors using the observed stellar mass-age and metallicity relation of nearby ETGs \citep{Thomas+05}.
As the high-redshift red galaxies form tight red sequence which is roughly consistent with the SSP models predicted from local red sequence,
the evolution of the color would be passive as discussed in previous studies \citep[e.g.,][]{Blakeslee+03a, Mei+06a, Mei+06b}.
Although our color selection limit, $i_{775}-z_{850} = 0.7$, is bluer than the modeled red sequence,
the selected galaxies should be quiescent
as the color limit indicates the SSP equivalent age of older than $\sim500$, $100$, and $300\mathrm{Myr}$ respectively at $z\sim1.0$, $1.2$, and $1.4$.

\subsubsection{Low-Redshift Quiescent Galaxies}
\begin{figure*}[htbp]
	\vspace{0\baselineskip}
	\centering
	\hspace{-2.0cm}	
	\includegraphics[width=11.0cm]{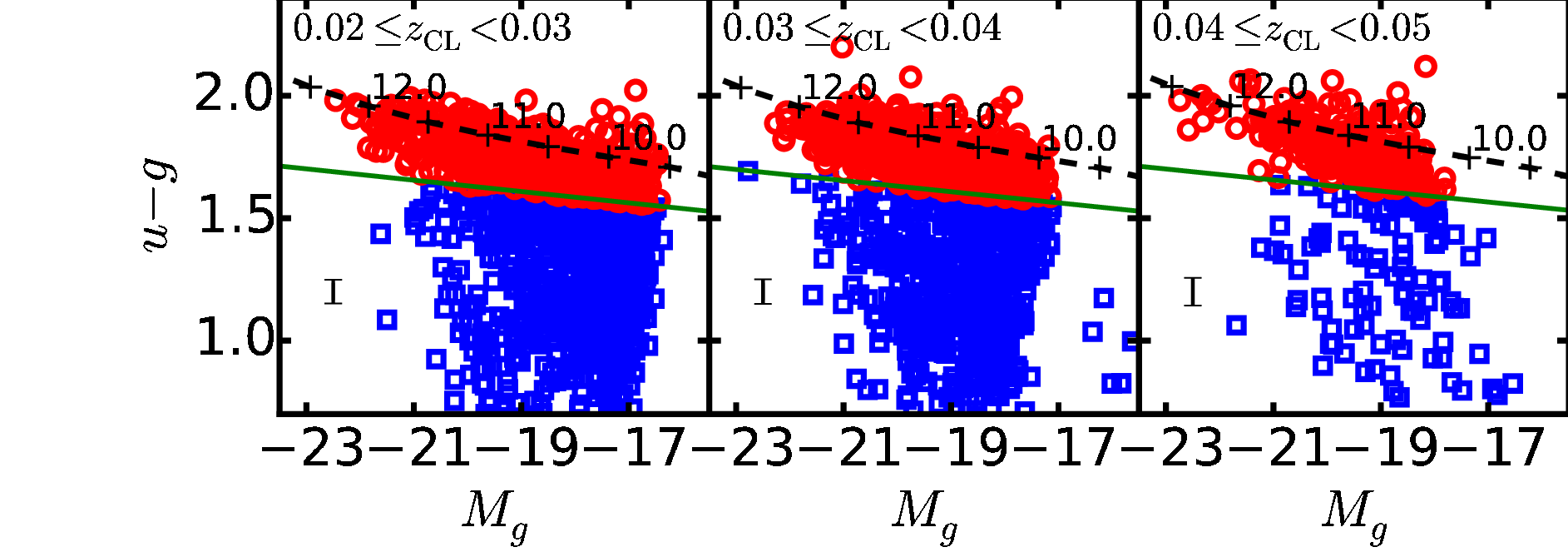}
	\caption{
		Color-magnitude diagrams of the low-redshift galaxies in three cluster redshift bins, 
		$0.02 \leq z_{CL} < 0.03$, $0.03 \leq z_{\mathrm{CL}} < 0.04$, $0.04 \leq z_{\mathrm{CL}} < 0.05$, from left to right.
		{\it Red circles} represent selected galaxies as quiescent while {\it blue squares} are unselected galaxies.
		Error bars indicate median uncertainties.
		{\it Dashed lines} represent the color-magnitude relation of the SSP model with the numbers at the ticks standing for the stellar mass, $\log(M_{*} / M_{\odot})$.
		{\it Green solid lines} are the separation lines above which a galaxy is regarded as quiescent.
		\label{fig: CMD_Loz}
	}
	\vspace{0\baselineskip}
\end{figure*}

We select the low-redshift quiescent galaxies using color-magnitude diagram of the $u-g$ color and absolute $g$ magnitude.
It is not a simple task to select possible descendants of the high-redshift quiescent galaxies 
as the luminosity and color of the galaxies evolve with redshift.
Since the evolution of each galaxy is unknown, we need to assume the evolution.
In this study, we assume that the high-redshift quiescent galaxies passively evolve with no mergers.
Although we simply select quenched galaxies for high redshift based on one single $i_{775}-z_{850}$ color,
we select  possible descendants at low redshifts based on color-magnitude diagram with more strict selection criteria
in order not to include newly quenched galaxies in $z<1$.
In Figure \ref{fig: CMD_Loz}, we plot the color-magnitude diagrams of the low-redshift galaxies in three cluster redshift bins.
We use the magnitude within the SDSS 3-arcsec fiber when we derive the $u-g$ color to reduce the effect of color gradients in a galaxy.
We refer to the Petrosian magnitude given in the SDSS catalog as the total $g$ magnitude.

In Figure \ref{fig: CMD_Loz}, we also plot the color magnitude relation by dashed lines inferred from the stellar mass-age and stellar mass-metallicity relation of nearby quiescent ETGs \citep{Thomas+05} using BC03 with Salpeter IMF.
Green solid lines in the figure are the bluer limit above which a galaxy is selected as quiescent.
We determined the bluer limit as follows.
We decrease the metallicity by three times the intrinsic scatter \citep[$\sigma_{[Z/H]}\sim0.08\mathrm{dex}$, see][]{Thomas+05} from the stellar mass-metallicity relation
and fix the stellar age to $7\mathrm{Gyr}$ which corresponds to the look-back time to $z\sim0.9$, the lowest redshift of the high-redshift galaxies.
Then, we derived the color-magnitude relation with these SSP parameters.
We finally fit a linear function the color-magnitude relation in the magnitude range of $-23 \leq M_{g} \leq -16$ to obtain the bluer limit.
As shown in Figure \ref{fig: CMD_Loz}, the bluer limit clearly separate red-sequence galaxies from those in the blue cloud.
The number galaxy in the low-redshift quiescent galaxy sample is 1733.

\subsection{Stellar Mass Limits}
\label{Subsec: StMassLim}
Next we impose a stellar mass limit.
We estimate a stellar mass of the galaxies by fitting SSP models of BC03 with Salpeter IMF to color(s) and magnitude.
For low-redshift galaxies, we fit the SSP SED to the Petrosian 
$g$ magnitude and SDSS 3-arcsec fiber colors of $u-g$, $g-r$, $g-i$, and $g-z$.
The free parameters are the stellar mass, age, metallicity.
One sigma uncertainty is calculated from errors 
in the $g$ magnitude and four colors via Monte-Carlo simulation.

For high-redshift galaxies, as only $i_{775}$ and $z_{850}$ magnitudes are available, 
only two free parameters can be constrained by the fitting.
As our high-redshift quiescent galaxies are largely consistent with red sequence
expected from local stellar mass-age and mass-metallicity relation (see Figure \ref{fig: CMD_Hiz}),
we relate the stellar age to mass using Equation (3) in \citet{Thomas+05}.
Then, independent fitting parameters are the stellar mass and metallicity.
We fit the synthetic SSP SED to the $z_{850}$ Petrosian magnitude and $i_{775}-z_{840}$ color measured in the central region ($0\farcs22$ diameter aperture) of a galaxy.
We also estimate the stellar mass by assuming the stellar mass-metallicity relation instead of the mass-age relation.
We adopt the absolute value of one half of the difference between stellar masses obtained in the two different ways
as uncertainty of the stellar mass for the high-redshift galaxies,
which is typically larger than one sigma error arising simply from uncertainties in $i_{775}$ and $z_{850}$ photometry.

With the estimated stellar masses,
we selected galaxies with $\log(M_{*}/M_{\odot}) \geq 10.5$ in order to make the stellar mass range comparable between low- and high-redshift samples.
For quiescent galaxies, the magnitude limit of the high-redshift sample is $\sim23-24$ mag in $z_{850}$ depending on clusters.
For the high-redshift quiescent sample, the stellar mass limit corresponds to the absolute magnitude of $M_{z_{850}}\sim-21.8$ 
and apparent magnitudes of $m_{z_{850}}\sim22.3$, $22.8$, and $23.2$ respectively at $z=1.0$, $1.2$, and $1.4$.
For the low-redshift quiescent galaxies, the stellar mass limit corresponds to $M_{g}\sim-19.5$ and $M_{r}\sim-20.3$ in $g$ and $r$ bands, respectively.
The latter becomes the $r$ band apparent magnitude of $m_{r}=16.4$ at $z=0.05$ (the upper redshift limit of our low-redshift sample)
which is brighter than the magnitude limit of the SDSS spectroscopic sample.
We have 158 and 513 galaxies for mass limited high- and low redshift quiescent samples.

We checked possible systematics between stellar masses obtained from one color and from multi colors.
First, we compared the stellar masses of low-redshift quiescent galaxies.
We computed stellar masses of low-redshift quiescent galaxies from $u-g$ fiber color and $g$ Petrosian magnitude assuming the mass-age relation \citep{Thomas+05}.
We have found that the stellar masses obtained from a single $u-g$ color tend to be over estimated by $\sim0.5$ dex.
Then, we compared the stellar masses of high-redshift quiescent galaxies.
For some of our high-redshift quiescent sample, \citet{Delaye+14} derived stellar masses from four-band photometries, $i_{775}$, $z_{850}$, $J$, and $K_{s}$.
We matched our high-redshift quiescent sample and Table B1 in \citet{Delaye+14},
and compared the stellar masses derived from a single color and three colors (four bands).
Unlike low-redshift samples, the stellar masses derived from a single $i_{775}-z_{850}$ color are on average under estimated by $\sim0.2$ dex.
Thus, we do not apply any correction to the stellar masses of our high-redshift galaxies.
We note that as \citet{Delaye+14} use \verb"MAG_AUTO" magnitudes (i.e., total magnitudes) in the four filters,
negative color gradients 
can not explain the difference in the stellar masses.
We would remind readers that the stellar masses of the high-redshift galaxies may have uncertainty of $\sim0.2$ dex.

\subsection{Selection of the Quiescent ETGs}
\label{Subsec: SelQETGs}
We then select ETGs from quiescent galaxies with morphological classification using photometric parameters.
We note that our conclusions, e.g., the disky-to-boxy fraction, are stable and have negligible change even if we use quiescent galaxy sample.

There are various ways to classify galaxy morphology.
One is the visual classification which has a long history in morphological classification (e.g, \citealp{Sandage61, Dressler80, Sandage+81}; recent studies by \citealp{Fukugita+07} for low-redshift with SDSS; \citealp{Postman+05} for high-redshift galaxy with {\it HST}).
There have also been classification using the concentration \citep{Morgan58},
and parameter combination of the concentration and mean surface brightness \citep{Doi+93, Abraham+94}, 
asymmetry \citep{Abraham+96}, or smoothness \citep{Conselice03, Yamauchi+05}.
Gini index is also adopted instead of the concentration parameter \citep{Abraham+03}.
Recently, machine learning scheme is introduced by \citet{Huertas-Company+11}.
In this study, we make use of the pair of the concentration parameter and mean surface brightness
which we have found less likely to be affected by signal-to-noise ratios of images than other parameters.
In Appendix, we provide results of simulations comparing the stability of the measurement of the Gini coefficient, asymmetry, concentration index, and mean surface brightness against signal-to-noise ratio.

Before the morphological classification, we run \verb"GALFIT" \citep{Peng+02} on cutout image of $z_{850}$ for high-redshift galaxies and $g$ for low-redshift galaxies
to fit a single S\'{e}rsic profile \citep[][for analytical properties of the profile]{Sersic+68, CiottiBertin99}
in order to derive some basic parameters of our galaxies and to create interloper-subtracted images.
We constrain the S\'{e}rsic index $n$ between 0.2 and 8.0,
and we input a PSF image to convolve with its model before fitting to the actual galaxy image.
We mask or fit simultaneously nearby objects depending on the degree of overlap.
For high-redshift galaxies, the PSF images and \verb"SExtractor" catalogs of nearby objects are constructed when we derive $i_{775}-z_{850}$ colors (see Subsection \ref{Subsubsec: Hiz QGs}).
For low-redshift galaxies, we also prepare PSF images and \verb"SExtractor" catalogs.
PSF images are created as for the high-redshift sample, by averaging images of non-saturated $\sim30$ stars in the original 2k$\times$4k frame of the target.
We make use of Cold/Hot method \citep{Rix+04} as high-redshift, when running \verb"SExtractor".
We obtained PSF deconvolved structural parameters such as S\'{e}rsic index $n$, effective radius $r_{e}$, axis ratio $q=b/a$, position angle, and surface brightness $\mu_{e}$ at $r_{e}$.
We do not use these structural parameters for morphological classification as not all galaxies can be successfully fitted by a single S\'{e}rsic profile.

\subsubsection{Morphological Classification with $C_{in}$ and $SB_{24.5}$}
The concentration index and mean surface brightness are measured in the similar manner as described in \citet{Doi+93}.
We first determine an isophote aperture by collecting pixels above cosmological dimming corrected surface brightness of 24.5 $\mathrm{mag\ arcsec^{-2}}$.
We use the smoothed images with a Gaussian kernel of $\sigma=2$ pixel to determine the isophote.
The mean surface brightness $SB_{24.5}$ is computed as the total flux within the aperture devided by the total area $A_{\mathrm{aper}}$.
We derive the equivalent outer radius as $r_{\mathrm{out}} = \sqrt{ A_{\mathrm{aper}} / \pi }$ and inner radius $r_{\mathrm{in}} = \alpha\ r_{\mathrm{out}}$,
where $\alpha$ is set to 0.3 in this paper.
The concentration index $C_{in}$ is defined as the ratio between the fluxes within a circular aperture with $r_{\mathrm{in}}$ and that with $r_{\mathrm{out}}$.

\citet{Doi+93} present that galaxies with S\'{e}rsic index of $n=4$ (ETGs) and $n=1$ (disk galaxies) can be separated in $C_{in}$-$SB_{24.5}$ plane.
However, ETGs with low surface brightness ($\mu_e \gtrsim 23-24 \mathrm{mag\ arcsec^{-2}}$) and disks with the brightness of $\mu_{\mathrm{e}} \gtrsim 23  \mathrm{mag\ arcsec^{-2}}$
overlap on the plane depending on the PSF size \citep[see Figures 1 and 2 in][]{Doi+93}.
Here, $\mu_{\mathrm{e}}$ is the surface brightness at effective radius $r_{e}$ (note that in \citealt{Doi+93}, $\mu_{\mathrm{e}}$ denotes the central brightness for $n=1$ galaxies).
As our galaxies reside in massive clusters, we have a certain portion of luminous ellipticals with low surface brightness \citep{Kormendy77},
and these galaxies will drop out from ETG classification if we simply apply the separation criteria described in \citet{Doi+93}.
Considering that ETGs and disk galaxies have different surface brightness-magnitude relation \citep[see for example Figure 20 in][]{Kormendy+12},
ETGs with $\mu_{\mathrm{e}} \gtrsim 23-24$ $\mathrm{mag\ arcsec^{-2}}$ and disks with $\mu_{\mathrm{e}} \gtrsim 23$ $\mathrm{mag\ arcsec^{-2}}$ 
would appear in different magnitude ranges.

\begin{figure*}[htbp]
	\vspace{0\baselineskip}
	\centering
	\hspace{-0.5cm}
	\centering 
 	\leavevmode 
 	\includegraphics[width=8cm]{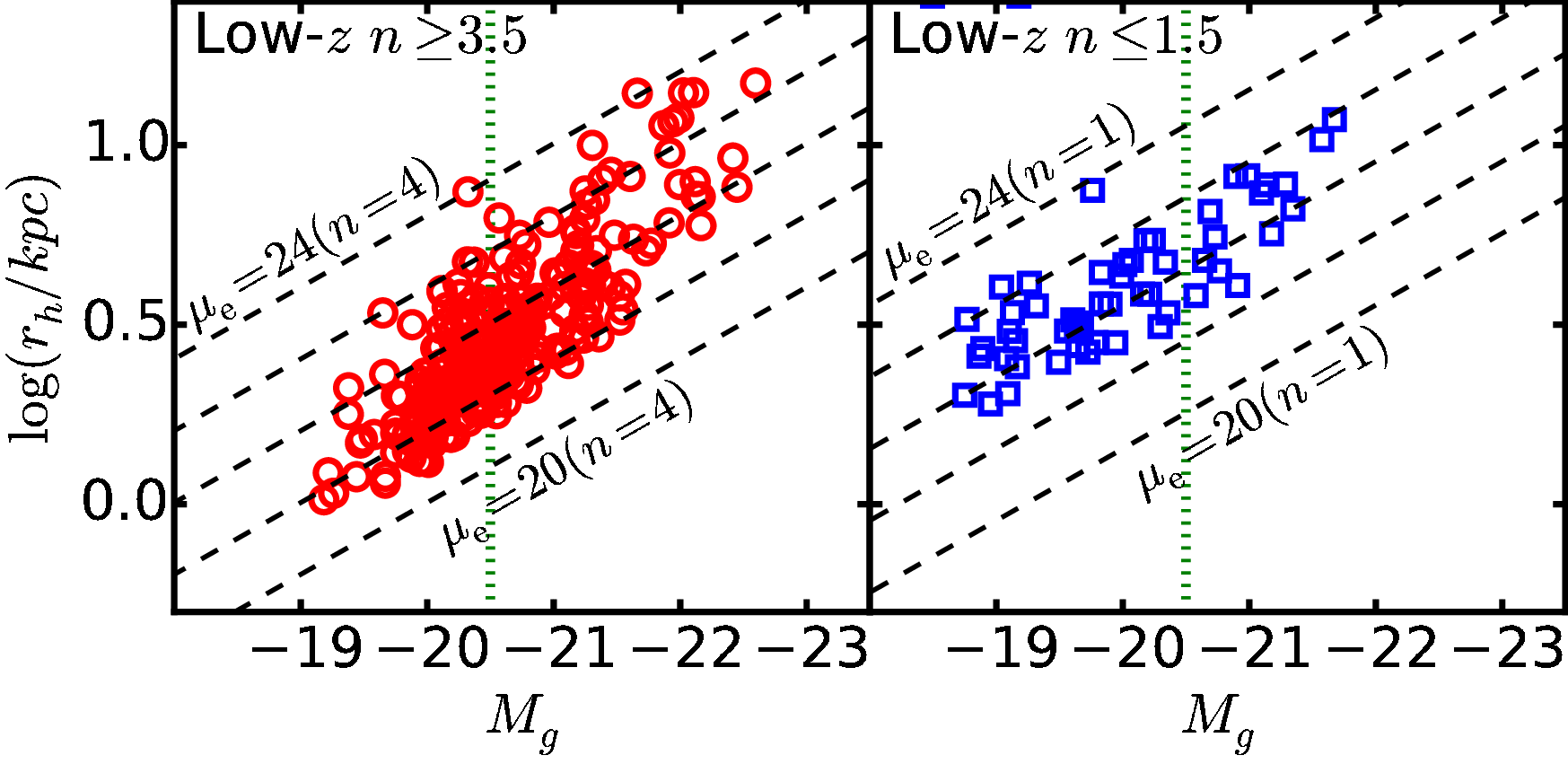}%
	\hfil 
	\includegraphics[width=8cm]{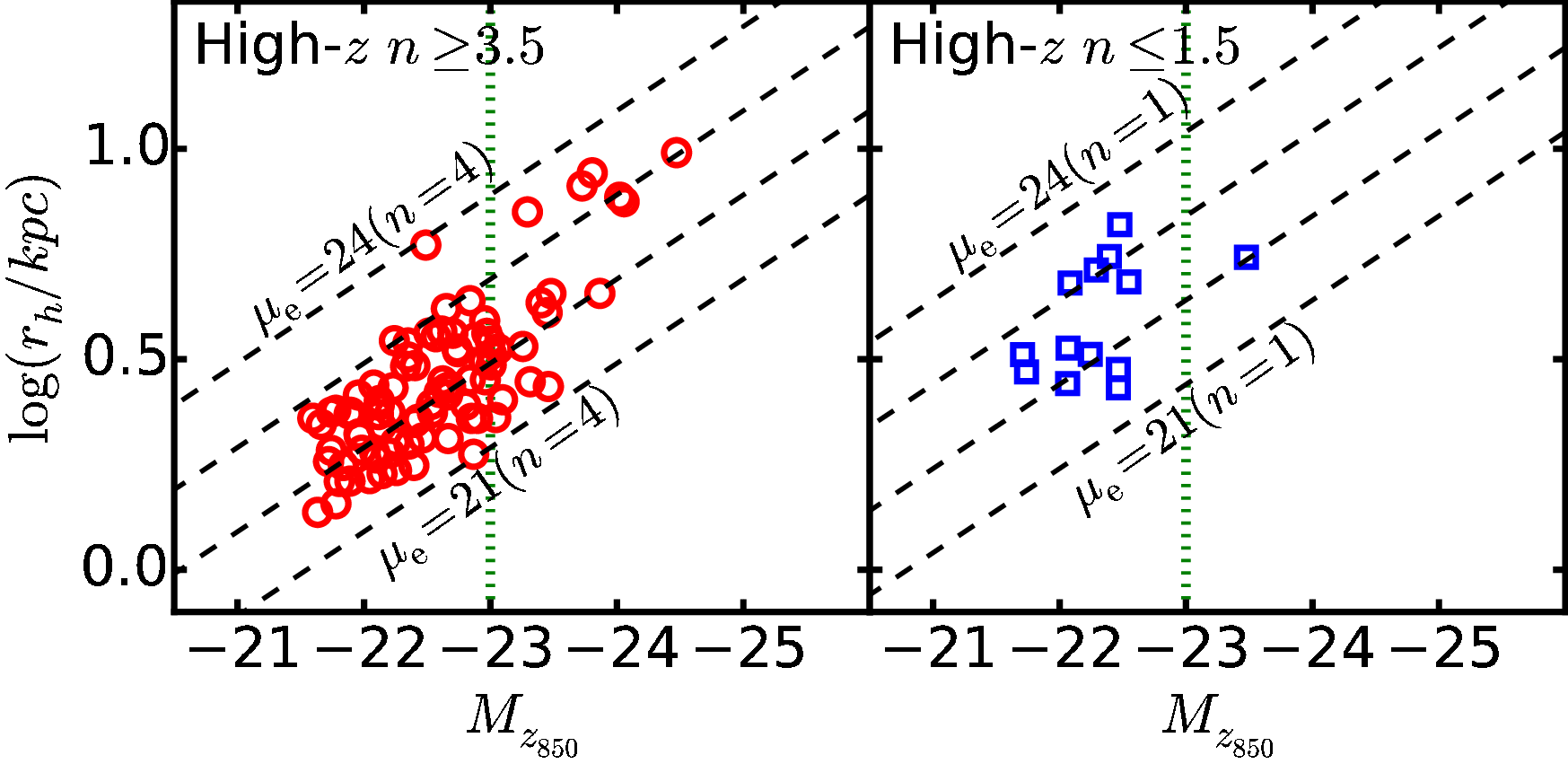}%
	\caption{
		{\bf Left}: Distribution of the low-redshift mass-limited quiescent galaxies with S\'{e}rsic indices of $n\geq3.5$ ({\it left panel})
		and $n\leq1.5$ ({\it right panel}) on the magnitude and half-light radius plane.
		{\it Dashed lines} indicate $M_{g}$-$r_{\mathrm{h}}$ relation assuming S\'{e}rsic profiles with $n=4$ ({\it left panel})
		and $n=1$ ({\it right panel}).
		$\mu_{\mathrm{e}}$ is set to 24, 23, 22, 21, 20$\mathrm{mag arcsec^{-2}}$ from top to bottom.
		\label{fig: Mg_Rh_Loz}
		{\bf Right}: Same as left but for high-redshift mass-limited quiescent galaxies.
		The magnitude is given as $M_{z_{850}}$ instead of $M_{g}$.
		For the model lines, $\mu_{\mathrm{e}}$ is set to 24, 23, 22, 21$\mathrm{mag arcsec^{-2}}$ from top to bottom.
		\label{fig: Mg_Rh_Hiz}
	}
	\vspace{0\baselineskip}
\end{figure*}

In Figure \ref{fig: Mg_Rh_Loz} (left), the distribution of the low-redshift quiescent mass-limited sample galaxies
with S\'{e}rsic indices of $n\geq3.5$ ({\it left panel}) and $n\leq1.5$ ({\it right panel})
are shown on the absolute $g$ magnitude and half-light radius plane.
The half-light radius is derived from \verb"FLUX_RADIUS" statistics obtained by \verb"SExtractor".
We also plot magnitude-radius relation inferred from S\'{e}rsic profile \citep[see][]{CiottiBertin99} with $n=4$ and $n=1$ with different $\mu_{\mathrm{e}}$.
The low-redshift galaxies with large $n$ and $\mu_{\mathrm{e}}\geq23-24 \mathrm{mag\ arcsec^{-2}}$ are brighter than $M_{g} \sim 20.5$, and there are very small number of low surface brightness ETGs in the fainter region.
At the same time, the galaxies with small $n$ and $\mu_{\mathrm{e}}\geq23 \mathrm{mag\ arcsec^{-2}}$ are fainter than $M_{g} \sim -20.5$.
Therefore, we define different selection criteria of ETGs on the $C_{in}$-$SB_{24.5}$ plane for the low-redshift galaxies brighter and fainter than $M_{g} = -20.5$ so that we can include luminous ETGs with low surface brightness simultaneously excluding disk galaxies.

Figure \ref{fig: Mg_Rh_Hiz} (right) is the distribution of the high-redshift quiescent mass-limited sample galaxies
on the absolute $z_{850}$ magnitude and half-light radius plane.
Disk-like galaxies with $n \leq 1.5$ with $\mu_{\mathrm{e}}\gtrsim23$ only appears in $M_{z_{850}}\gtrsim-23$,
and luminous ETGs with $\mu_{\mathrm{e}}\gtrsim23$ in  $M_{z_{850}}\lesssim-23$.
Thus, we set the critical magnitude to $M_{z_{850}}=-23.0$
above or below which the election criteria of ETGs on the $C_{in}$-$SB_{24.5}$ plane are defined differently.
\begin{figure*}[htbp]
	\vspace{0\baselineskip}
	\hspace{-0.5cm}
	\centering 
 	\leavevmode 
 	\includegraphics[width=8cm]{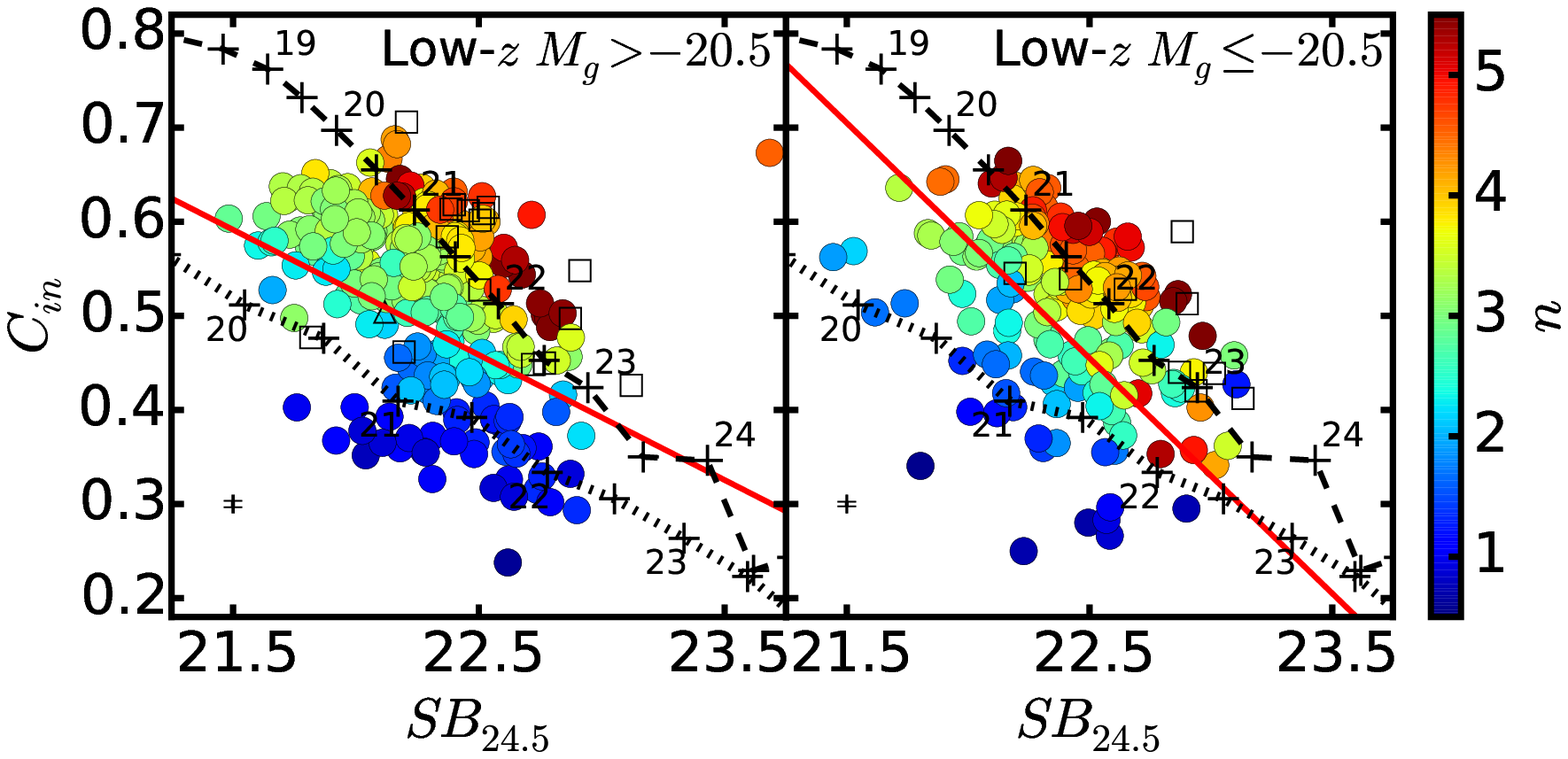}%
	\hfil 
	\includegraphics[width=8cm]{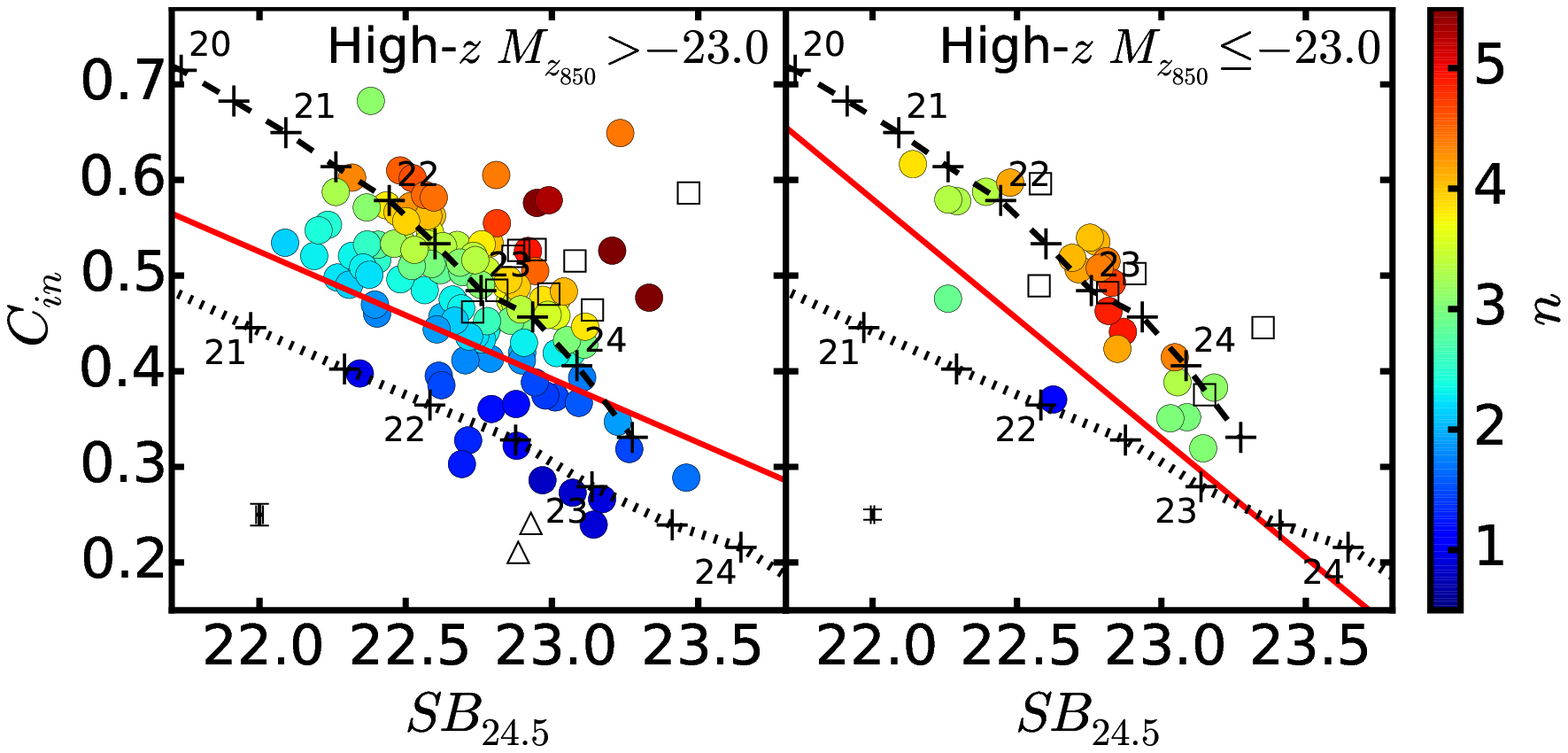}%
	\caption{
		{\bf Left}: Distribution of the low-redshift mass-limited quiescent galaxies with $M_{g} > -20.5$ ({\it left panel}) and $M_{g} \leq -20.5$ ({\it right panel})
		on the $C_{\mathrm{in}}$-$SB_{24.5}$ plane.
		Color code represents the S\'{e}rsic index $n$.
		{\it Open squares} and {\it open triangles} indicates galaxies with
		$n=8$ and $0.2$, respectively.
		{\it Dashed} and {\it dotted line} show expected positions for galaxies with S\'{e}rsic profiles with $n=4$ and $1$, respectively.
		The number on the ticks indicates the surface brightness $\mu_{\mathrm{e}}$.
		{\it Red solid line} is the separation line used for ETG selection.
		\label{fig: CinSB_Loz}
		{\bf Right}: Same as left but for high-redshift mass-limited quiescent galaxies with the separating magnitude of $M_{z_{850}}=-23.0$.
		The surface brightness $SB_{24.5}$ is corrected for the passive evolution.
		\label{fig: CinSB_Hiz}
	}
	\vspace{0\baselineskip}
\end{figure*}

In Figure \ref{fig: CinSB_Loz}, the distributions of the low- and high-redshift quiescent sample on the $C_{\mathrm{in}}-SB_{24.5}$ plane are shown.
We also plot the expected positions for model galaxies with $n=4$ ({\it dashed lines}) and $n=1$ ({\it dotted lines}) S\'{e}rsic profile convolved with typical PSF.
We selected galaxies above the separation line ({\it red solid line}) as ETGs.
S\'{e}rsic index is shown by the color code.
As one can see in the figure, galaxies with $n\gtrsim2$ are selected as ETGs.
The number of galaxy with $\log(M_{*}/M_{\odot}\geq10.5)$ in the low- and high-redshift quiescent ETG sample are 355 and 130, respectively.
Note that majority of the low- and high-redshift quiescent galaxy is ETGs,
and this ETG selection hardly affects the result.

\subsection{Basic Properties of the Quiescent ETG Samples}
\label{BP_Q_ETGs}
We present the basic properties of the quiescent ETG samples
such as the stellar mass, size, axis ratio, and S\'{e}rsic index
measured with \verb"GALFIT".
Hereafter, when we discuss the structural parameters obtained with \verb"GALFIT", we exclude galaxies fitted with $n=$0.2 and 8.0 as these values are limit of the parameter constraints and may not be reliable.
There are 14 and 23 objects with $n=$2.0 or 8.0 respectively in the high- and low-redshift quiescent ETG samples.

\begin{figure*}[htbp]
	\vspace{0\baselineskip}
	\centering
	\hspace{-0.5cm}	
	\includegraphics[width=12.0cm]{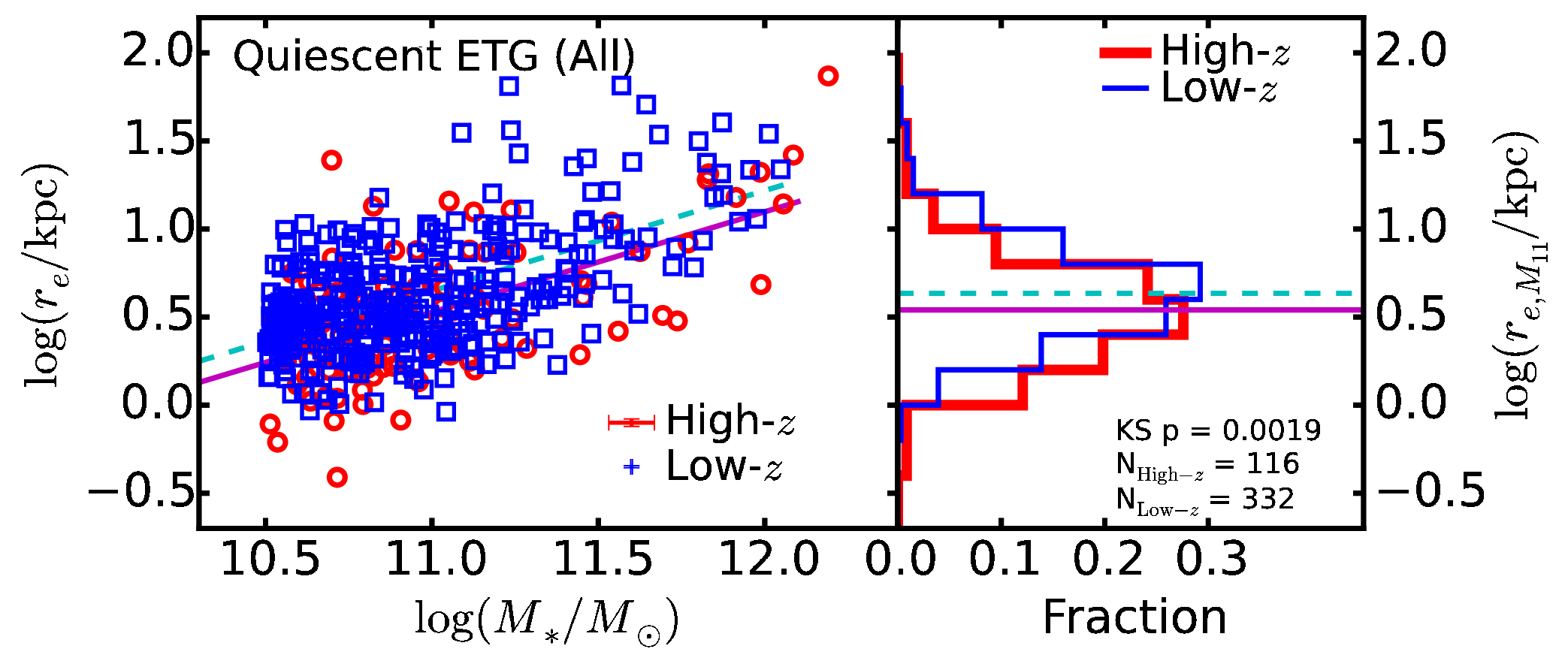}
	\vspace{0\baselineskip}
	\caption{
		Mass-size relation of the low- ({\it blue squres}) and high-redshift {\it red circles}) quiescent ETGs
		is shown in the left panel.
		The median errors are shown in the panel.
		{\it Cyan dashed} and {\it magenta solid} lines indicate linear function fitted to the relation for low- and high-redshift samples.
		In the right panel, histograms of the mass normalized size $r_{\mathrm{e}, M_{11}}$ are plotted.
		{\it Cyan dashed} and {\it magenta solid} lines indicate
		the intercept of the fitting lines and $\log(M_{*}/M_{\odot})=11$ in the left panel
		for low- and high-redshift samples, respectively.
		\label{fig: M_Re_all}
	}
	\vspace{0\baselineskip}
\end{figure*}

In the left panel of Figure \ref{fig: M_Re_all},
the effective radii of the low- and high-redshift quiescent ETGs are plotted against the stellar mass (the mass-size relation).
We fit a linear function $\log(r_{\mathrm{e}}/\mathrm{kpc}) = a \times ( \log(M_{*}/M_{\odot}) - 11) + b$.
We fix $a$ to 0.57 following \citet{Delaye+14} and only $b$ is a free parameter.
The fitted lines are shown in cyan dashed and magenta solid lines for the low- and high-redshift samples, respectively.
We also calculate the mass normalized size $r_{\mathrm{e}, M_{11}}$
which is obtained as $\log(r_{\mathrm{e}, M_{11}}) = \log(r_{\mathrm{e}}/\mathrm{kpc}) - 0.57  \times ( \log(M_{*}/M_{\odot}) - 11) $.
In the right panel of Figure \ref{fig: M_Re_all}, the histograms of $\log(r_{\mathrm{e}, M_{11}})$ are shown for low- and high-redshift samples.
The median mass normalized sizes are $<\log(r_{\mathrm{e}, M_{11}})> = 0.63 \pm 0.02$ and $0.54\pm 0.03$ for the low- and high-redshift samples, respectively.
The overall distribution shifts towards the larger size from $z\sim1$ to $0$.
Kolmogorov-Smirnov (KS) test gives the p-value of 0.001 which indicates the two samples are statistically different.
The size evolution is basically consistent with the result presented in a previous study \citep{Delaye+14}.

\begin{figure*}[htbp]
	\vspace{0\baselineskip}
	\centering
	\hspace{-0.5cm}	
	\includegraphics[width=11.0cm]{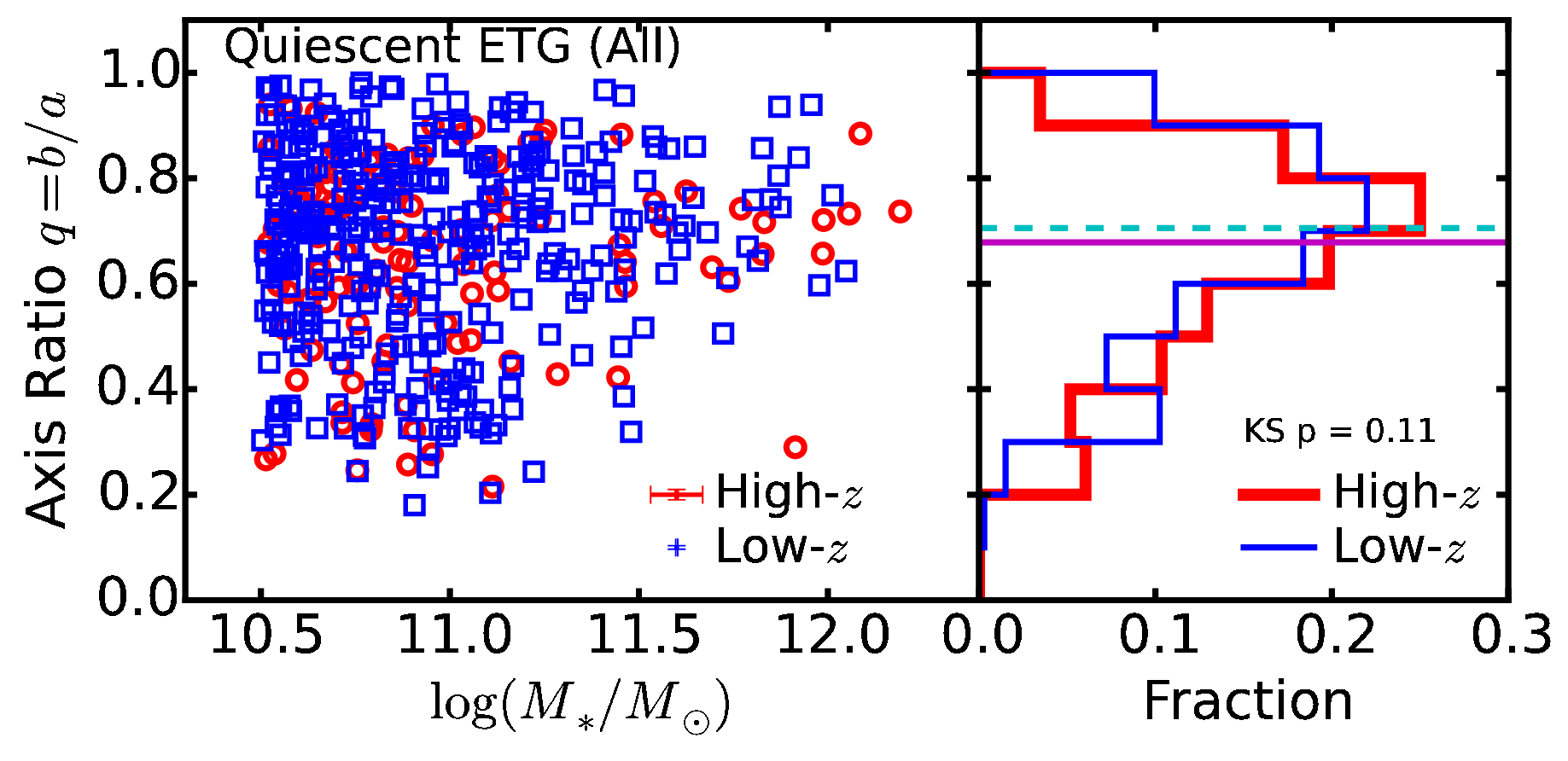}
	\vspace{0\baselineskip}
	\caption{
		Axis ratios of the low- ({\it blue squres}) and high-redshift {\it red circles}) quiescent ETGs
		is plotted against the stellar mass in the left panel.
		The median errors are shown in the panel.
		In the right panel, histograms of the axis ratios are plotted.
		{\it Cyan dashed} and {\it magenta solid} lines indicate the median value for low- and high-redshift samples, respectively.
		\label{fig: M_q_all}
	}
	\vspace{0\baselineskip}
\end{figure*}

In Figure \ref{fig: M_q_all},
the axis ratios are plotted against the stellar mass in the left panel,
and the histograms of the axis ratios are plotted in the right.
The median axis ratios are $<q> = 0.71 \pm 0.01$ and $0.68 \pm 0.02$ for the low- and high-redshift samples, respectively.
The median axis ratio of the high-redshift sample is smaller than that of the low-redshift one,
which is consistent with a previous work \citep{De_Propris+15},
but in this study, the KS p-value is not enough small ($p=0.11$) to conclude that the two distributions are statistically different.

\begin{figure*}[htbp]
	\vspace{0\baselineskip}
	\centering
	\hspace{-0.5cm}	
	\includegraphics[width=11.0cm]{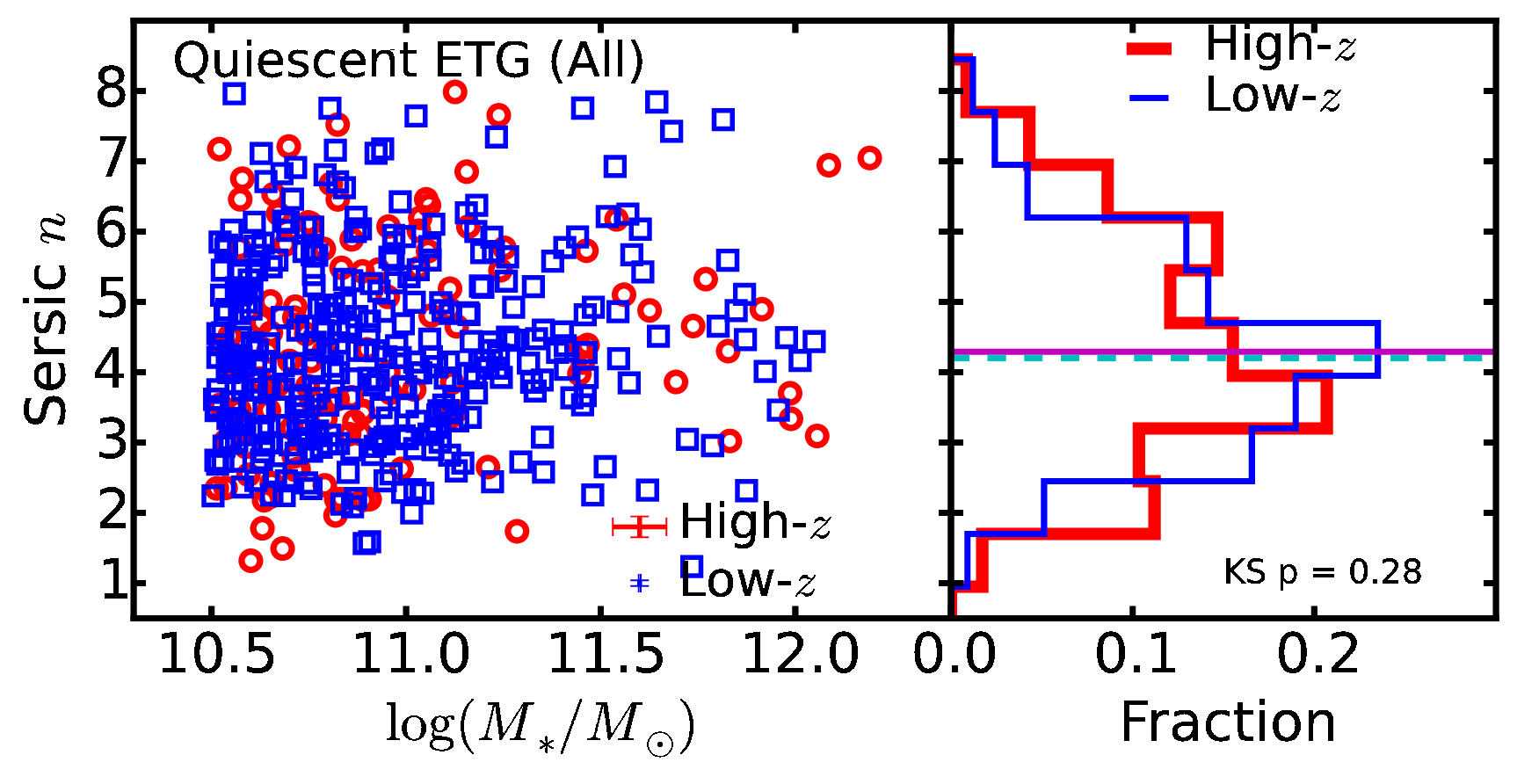}
	\vspace{0\baselineskip}
	\caption{
		S\'{e}sic index of the low- ({\it blue squres}) and high-redshift {\it red circles}) quiescent ETGs is shown in the left panel.
		The median errors are shown in the panel.
		In the right panel, histograms of the S\'{e}sic index are plotted.
		{\it Cyan dashed} and {\it magenta solid} lines indicate the median value for low- and high-redshift samples, respectively.
		\label{fig: M_n_all}
	}
	\vspace{0\baselineskip}
\end{figure*}

In Figure \ref{fig: M_n_all},
the S\'{e}sic indices are plotted against the stellar mass in the left panel,
and the histograms of the S\'{e}sic indices are plotted in the right panel.
The median  S\'{e}sic indices are $<n> = 4.3 \pm 0.01$ and $4.2 \pm 0.2$ for the low- and high-redshift samples, respectively.
The median values are consistent within uncertainty.
The KS test gives the p-value of 0.28,
which indicates that the two samples could be drawn from the same sample.



\section{Measuring Isophote Shapes}
\label{Sec: method}
We developed an isophote analysis code which is optimized for high-redshift galaxies with low surface brightness and small angular size,
based on \citet{Bender+87}.
The code takes three steps:
contour determination;
deviated ellipse fit;
and estimation of errors.
Readers who are not interested in the detail of the method,
please skip this section.

\subsection{Contour Determination}
\label{subsec: Contour Determination}
In our code, the number of pixels needed to determine a contour is adjusted adaptively according to $S/N$ per pixel.
This point is different from other isophote analyzing code such as {\sc iraf} task {\sc ellipse} \citep{Jedrzejewski87}.
There are three main challenges in investigating isophote contours of high-redshift galaxies.
First, apparent surface brightness of high-redshift galaxy decreases with redshift as $(1+z)^{-4}$ (cosmological surface brightness dimming).
For $z\sim1$ galaxies, the surface brightness becomes dimmer by $\sim1/16$
although intrinsic luminosity of a galaxy becomes brighter at high redshift due to passive evolution as described in the previous section.
Second, small apparent angular size makes isophote shape measurement difficult
in that the number of pixels used to determine a contour becomes small.
The precision of a contour (or sampling points of the contour, $x_i, y_i$)
is enhanced by square root of the number of pixel used to determine the contour for a given surface brightness.
As easily imagined, 
for large, nearby galaxies, low-order isophote shape parameter, e.g., $a_4$, is rather insensitive to noise per pixel
since the typical scale length of $a_{4}$, $\sim r \times \pi / 4 $, is much larger than pixel size.
On the other hand, for distant galaxies with small apparent size, the typical scale length is close to the pixel size,
so noise per pixel affects the isophote shape measurement more severely.
In addition, the large PSF size compared to the apparent galaxy size may introduce systematic errors, which will be discussed in Subsection \ref{subsec: eff_im_qual}.

First of all, pixels used in the isophote shape analysis is selected in the following way.
Pixels with a flux above a given detection limit is picked up using a smoothed image with Gaussian kernel of $\sigma=1$pixel.
In this paper, we set the detection limit one sigma background noise, $\sigma_{\mathrm{bkg}}$, above background level.
Then, from the selected pixels, those contiguous to the initial center of the target are chosen.
The position of the initial center is given as an input.

Bright objects close to the target are masked.
We make use of the output from {\sc galfit}.
As objects close to the target are simultaneously fitted,
we mask the pixel where the modeled flux of nearby objects exceeds that of the target object.
We have confirmed that bright nearby objects are successfully masked even in the central region of a galaxy cluster in this way.

We extract pixel annuli from the selected pixels,
and isophote contours are sampled from the annuli.
In the first step, center of the target is identified, and pixels which are likely to be affected by PSF are discarded.
The center of a galaxy is identified as the intensity peak within the brightest 10$\%$ of the pixels.
We note that our target of interest is ETGs whose light is concentrated, and flux peak of a galaxy is not severely affected by noise.

The, pixels which may be affected strongly by PSF are discarded in the following way.
The faintest pixel within a PSF radius $r_{\mathrm{PSF}}$ from the center is identified,
and pixels within an isophote of the intensity of the faintest pixel are masked and discarded.
We refer these discarded pixels as ($pix_{\mathrm{dis}}$) hereafter.
In this study, we set $r_{\mathrm{PSF}}$ to the PSF HWHM,
i.e., 1.0 pix for {\it HST} images and 1.64 pix for SDSS.

In the next step, the first annulus (inner most annulus) is determined.
The pixels surrounding ($pix_{\mathrm{dis}}$) are the inner pixels of the first annulus.
We refer these inner pixels as ($pix_{1,  \mathrm{in}}$).
Then, of the pixels surrounding ($pix_{1, \mathrm{in}}$), the faintest pixel and its intensity are found,
and pixels contiguous to the center above the intensity are noted as ($pix_{1, \mathrm{out}}$).
The first annulus, ($pix_{1, ann}$), is defined by the pixels, 
\begin{eqnarray}
	(pix_{1, ann}) = (pix_{1, \mathrm{out}}) - (pix_{\mathrm{dis}}).
\end{eqnarray}

\begin{figure}[htbp]
	\plotone{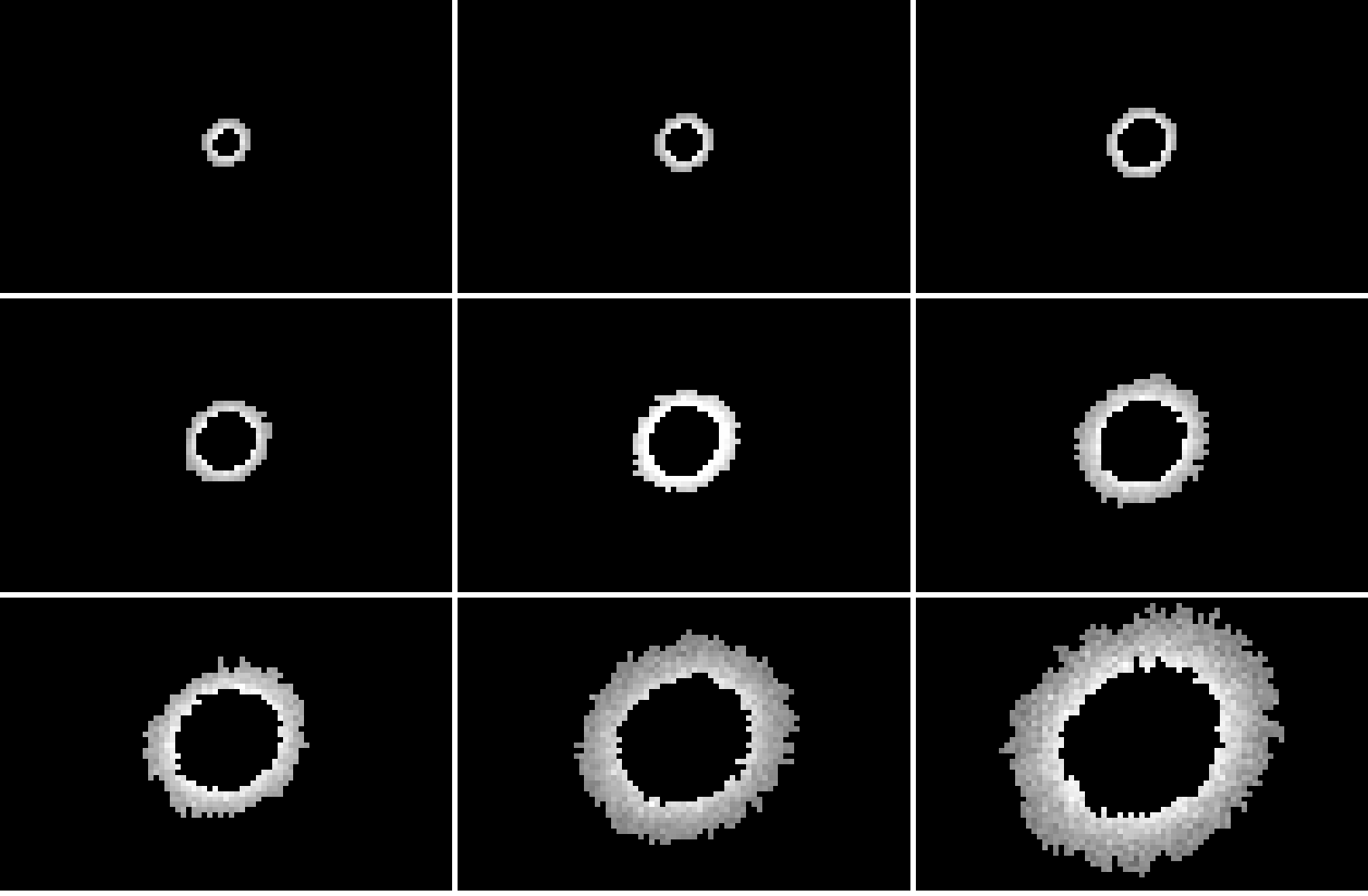}
	\caption{
		Example of determined successive annuli.
		The annuli are narrow near the center, but are wide in outer, low-$S/N$ region.
		\label{fig: Seccessive_Annuli}
	}
\end{figure}

The successive annuli are determined in the following way.
The inner pixels of $n$-th annulus is defined as 
\begin{eqnarray}
	(pix_{n, \mathrm{in}}) = (pix_{n-1, \mathrm{ann}}) - (pix_{n-1, \mathrm{in}}) \ \ (n \geq 2).
\end{eqnarray}
Then, of the pixels surrounding ($pix_{n, \mathrm{in}}$), the faintest pixel and its intensity are found,
and pixels contiguous to the center above the intensity are noted as ($pix_{n, \mathrm{out}}$).
The $n$-th annulus, ($pix_{n, \mathrm{ann}}$), is defined as
\begin{eqnarray}
	(pix_{n, \mathrm{ann}}) = (pix_{n, \mathrm{out}}) - (pix_{n-1, \mathrm{out}}) + (pix_{n, \mathrm{in}}).
\end{eqnarray}
Repeating this process until all the pixels defined in the previous subsection are used,
successive pixel annuli are determined as described in Figure \ref{fig: Seccessive_Annuli}.
The annuli tend to be narrow in high-S/N region, i.e., in the central region, but to be wide in low-S/N region, i.e., in the outskirt, 
As a contour is determined with larger number of pixel,
the sampling points of the contour are not significantly affected by noise even in a low-S/N region.

After pixel annuli are determined, the data points of contours are sampled.
First, the isophote level of the $n$-th contour is calculated as the weighted mean
of the pixels in the $n$-th annulus, 
\begin{eqnarray}
	I_{n} = \frac{ \sum_{\mathrm{ann}} { I_{x,y} /  \sigma_{I_{x,y}}^{2}  }} { \sum_{\mathrm{ann}}{1 /  \sigma_{I_{x,y}}^{2} } },
\end{eqnarray}
where $I_{x,y}$ and $ \sigma_{I_{x,y}}$ are the intensity and noise per pixel at $(x, y)$, respectively.

\begin{figure}[htbp]
	\hspace{0cm}	
	\plotone{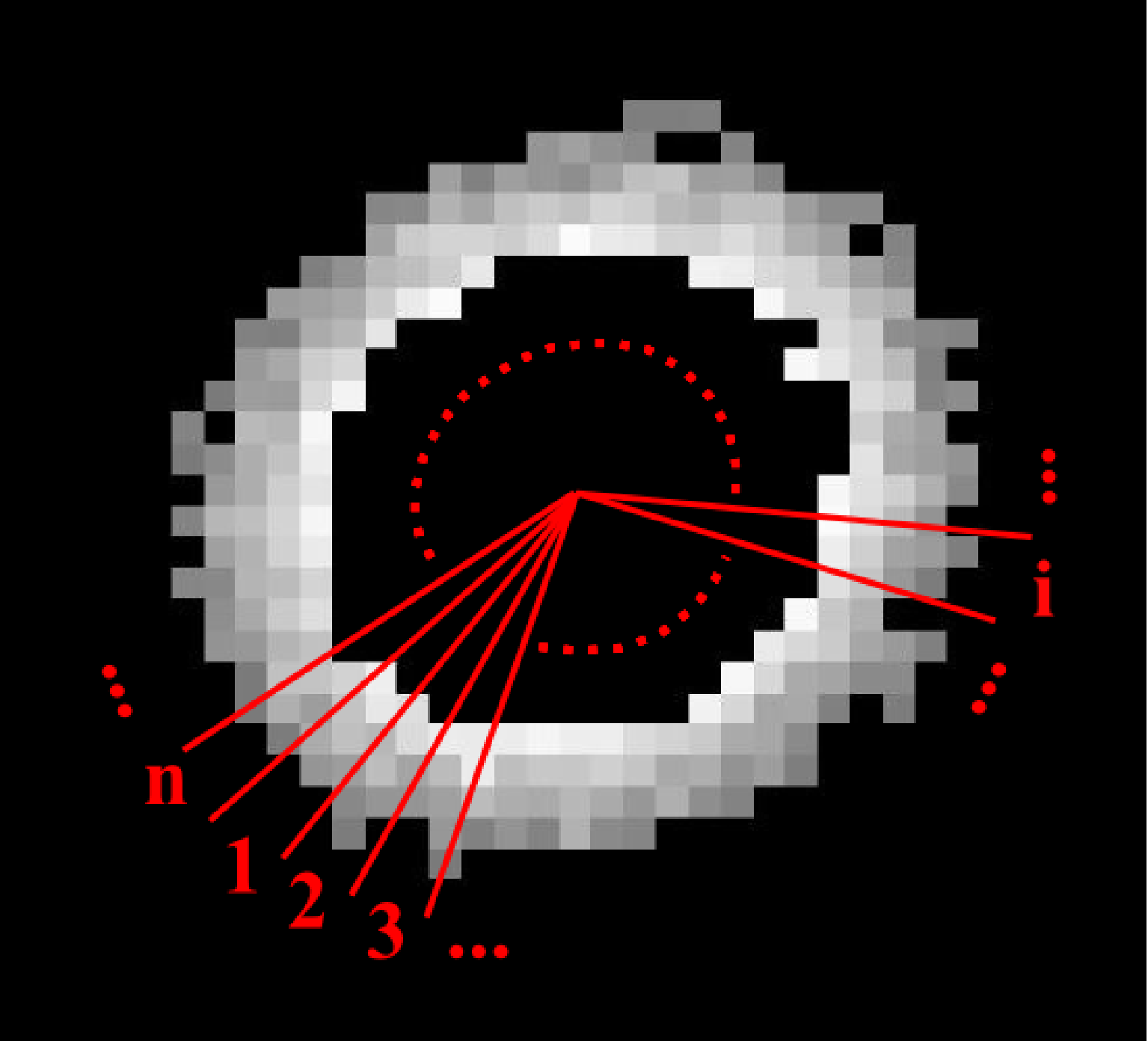}
	\caption{
		Example of an annulus divided into azimuthal bin.
		\label{fig: Azimuth_Bin}
	}
\end{figure}

Then, each annulus is divided into azimuthal bins, as described in Figure \ref{fig: Azimuth_Bin}.
The number of bin is three times the number of pixels in annulus,
or if it is larger than 90, the number of bin is set to 90.
Dividing lines are given so that the ellipse parameter $t_i$ of the lines becomes equidistant.
The ellipse parameter $t$ is a parameter appearing in the parametric formalization of an ellipse, 
\begin{eqnarray}
	x &=& a \cos{(t - \psi)}	\nonumber \\
	y &=& b \sin{(t - \psi )},
\end{eqnarray}
where $a$ and $b$ are the semi-major and minor axis, and $\psi$ is the position angle.
To define the dividing lines, the position angle, $\psi$, and axis ratio, $q = b/a$, are necessary before analyzing isophote shapes.
In this step, the position angle, $\psi$, and axis ratio, $q = b/a$, are fixed to the value
estimated from the intensity-weighted second-order moments
\citep[e.g.,][]{Stobie80, Lupton+01, Yamauchi+05}
within the brightest 25$\%$ of the pixels above one sigma isophote with nearby objects masked.

In each azimuthal bin, the sampling points of the isophote contour are determined using the local radial profile.
First, radius of each pixel in each bin is derived.
The radius of the pixel at $(x, y)$ is calculated as
\begin{eqnarray}
	r_{x, y} = \sqrt{(x - x_0)^2 + (y - y_0)^2},
\end{eqnarray}
where $(x_0, y_0)$ is the center of the target.
The error of intensity is scaled by the included area into the bin as
\begin{eqnarray}
	\sigma_{I_{x,y}}^{\prime 2}  = \frac{ \sigma_{I_{x,y}}^{2} }{A_{x, y, i}},
\end{eqnarray}
where
$\sigma_{I_{x,y}}$ is the intensity error per pixel, 
and $A_{x, y, i}$ ($\leq1$) is the included area into $i$-th azimuthal bin.
This scaling allows us to treat the errors as independent,
since $A_{x, y, i}$ is equivalent to the degree of freedom per pixel included into $i$-th bin.

We fit a linear function, $I_{\mathrm{fit}}(r) = \alpha\  r + \beta$, to the local radial profile $(r, I)$ using all pixels in each bin
by minimizing
\begin{eqnarray}
	\chi^2
	=
	\frac{
		\left( I_{x, y} - I_{\mathrm{fit}}(r) \right)^2
	}{
		\sigma_{I_{x,y}}^{\prime 2}
	},
\end{eqnarray}
and derive $\alpha$, $\beta$, and the covariance matrix from the scaled intensity errors $\sigma_{I_{x,y}}^{\prime}$.
Although a radial intensity profile of a galaxy is often described by non-linear function such as S\'{e}rsic function \citep{Sersic+68},
the radial range is enough small
for the local profile to be fitted by a linear function.
We have tested log intensity instead of intensity itself in this process,
but we have not found significant differences.

Then, the radius of sampling point $r_{n, i}$ is derived as the crossing point of the fitting line $I=I_{\mathrm{fit}}(r)$
and the isophote level $I=I_{n}$.
The error of the radius $\sigma_{r_{n, i}}$ is calculated from the covariance matrix,
i.e., the propagation of errors from $\sigma_{I_{x,y}}^{\prime}$ to $\sigma_{r_{n, i}}$ is calculated.
The $(x, y)$ position of the sampling point is then given as,
\begin{eqnarray}
	x_{n, i} &=& r_{n, i} \cos(\theta_{i} - \phi) + x_0	\nonumber \\
	y_{n, i} &=& r_{n, i} \sin(\theta_{i} - \phi) + y_0,
\end{eqnarray}
and errors as
\begin{eqnarray}
	\sigma_{x_{n, i}} &=& \sigma_{r_{n, i}} \cos(\theta_{i} - \phi)	\nonumber \\
	\sigma_{y_{n, i}} &=& \sigma_{r_{n, i}} \sin(\theta_{i} - \phi),
\end{eqnarray}
where $\theta_{i} = (t_i + t_{i+1}) / 2$.
Repeating this process for all azimuthal bins gives the sampling points for the $n$-th isophote contour.

\subsection{Deviated Ellipse Fit}
To each isophote contour, an ellipse is fitted with third- to sixth-order Fourier deviations.
The fitting parameters are
five parameters related to ellipse,
center $(x_0, y_0)$,
axis ratio ($q$),
position angle ($\psi$),
and semi-major axis ($a$),
and parameters related to the deviation, $a_n$ and $b_n$ $(n = 3,4,5,6)$.
The zero-th to the second-order deviation terms are not included,
since they degenerate with center, axis ratio, and position angle.
First, the initial values of the five ellipse parameters are derived by minimizing
\begin{eqnarray}
	\chi^2
	=
	\sum_{i}
	\left(
		(x_i - x_{e, i})^2
		+
		(y_i - y_{e, i})^2
	\right),
\end{eqnarray}
where $(x_i, y_i)$ is the sampling point of the contour, 
$(x_{e, i}, y_{e, i})$ is the point on the fitting ellipse given as
\begin{eqnarray}
	x_{e, i} &=& a \cos{(\theta_{i} - \psi)}		+ x_0 \nonumber \\
	y_{e, i} &=& b \sin{(\theta_{i} - \psi )}	+ y_0.
\end{eqnarray}
In this step, we iteratively fit the center, axis ratio, and position angle around the value determined in the previous subsection, 
whereas the the semi-major axis is fitted around mean value of $r_i$.

After initial conditions are determined, a deviated ellipse is fitted.
In this process, $(x_{e, i}, y_{e, i})$ is given as
\begin{eqnarray}
	x_{e, i} &=& ( a + \Delta r_{i} ) \cos{(\theta_{i} - \psi)}	+ x_0  \nonumber \\
	y_{e, i} &=& ( b + \Delta r_{i} ) \sin{(\theta_{i} - \psi )}	+ y_0  ,
\end{eqnarray}
where $\Delta r_{i}$ is the deviation term,
\begin{eqnarray}
	\Delta r_{i} = \sum_{k = 3}^{6} ( a_k \cos( k \theta_{i} - \psi ) + b_k \sin( k \theta_{i} - \psi ) ).
\end{eqnarray}
Finally, zero-th to sixth-order deviation terms are derived with ellipse parameters, $(x_0, y_0)$, $q$, $\phi$, and $a$, fixed to the values obtained above.

\begin{figure}[htbp]
	\vspace{0\baselineskip}
	\hspace{0cm}	
	\plotone{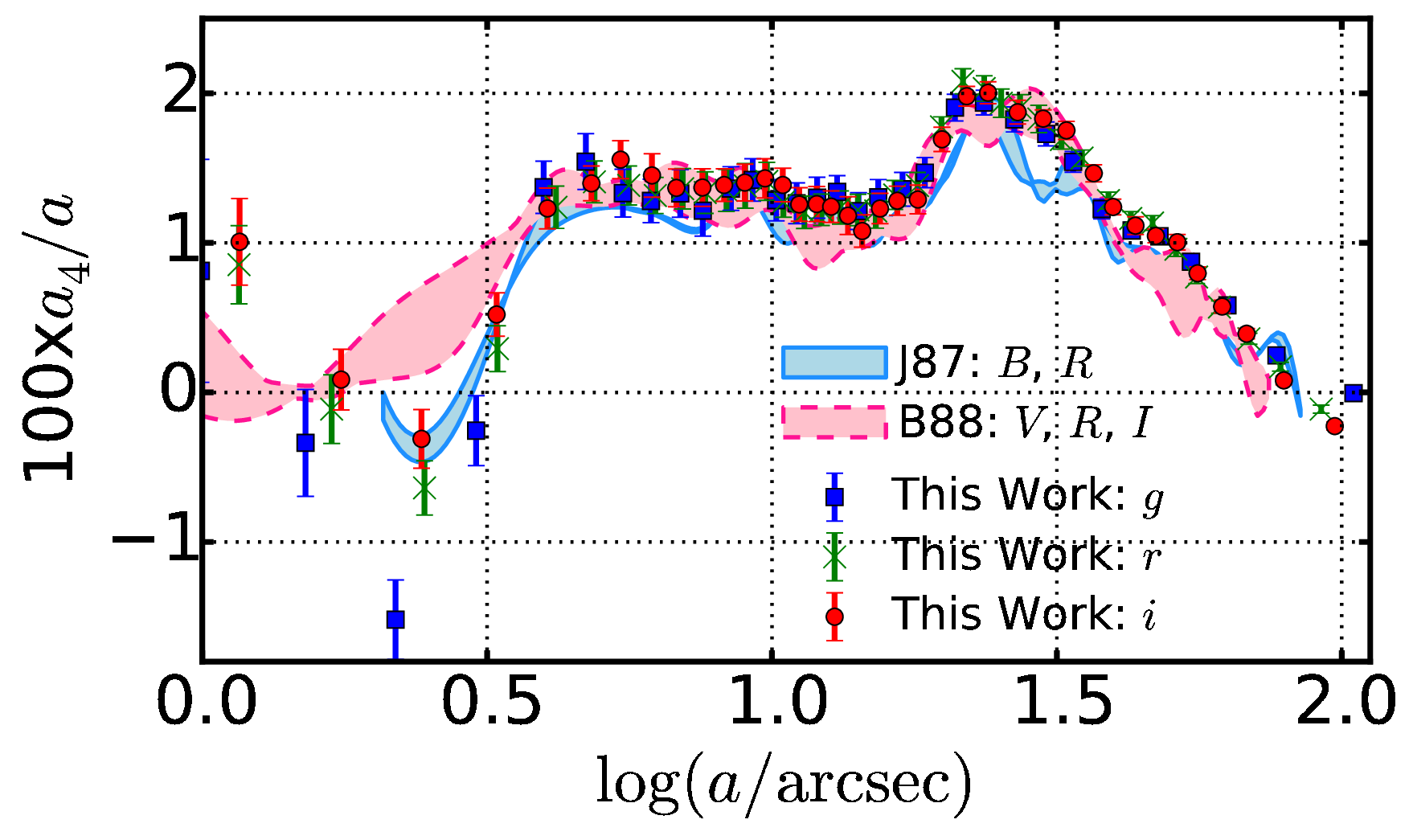}
	\caption{
		Example of radial (semi-major axis) profiles of the $a_{4}$ parameter of a nearby galaxy, NGC4697.
		{\it Blue shaded region enclosed by solid line} represents the result obtained by \citet[][in $B$ and $R$ band]{Jedrzejewski87}
		whereas {\it red shaded region enclosed by dashed line} indicates the measurements by \citet[][in $V$, $R$ and $I$ band]{Bender+88}.
		{\it Blue squares} are our measurements in $g$ band, {\it green crosses} are in $r$, and {\it red circles} are in $i$.
		\label{fig: NGC4697_isophote}
	}
	\vspace{0\baselineskip}
\end{figure}

In Figure \ref{fig: NGC4697_isophote}, an example of measurements of the $a_{4}/a$ parameters of a nearby galaxy NGC4697 is shown.
As isophote shapes of this galaxy have been investigated in \citet{Jedrzejewski87} and \citet{Bender+88},
we compare our $a_{4}$ measurements with these previous studies.
We measure the isophote shapes using SDSS $g$, $r$, and $i$ bands.
Our measurements are in good agreement with the previous studies except for the inner most region
where the seeing affects the measurement of $a_{4}$ and three measurements diverge probably due to the difference of the seeing.
The typical seeing FWHM in our measurement is $\sim1.1-1.3\rm{arcsec}$ estimated from unsaturated stars
while it amounts to $\sim1.5-1.9\rm{arcsec}$ in \citet{Jedrzejewski87} and $\sim2\rm{arcsec}$ in \citet{Bender+88}.

We reduce the radial profiles of various isophote shape parameters such as $q$, $\phi$, $a_{0-6}$, $b_{0-6}$ to mean values in the following way
as it is too complicated to compare the radial profile between galaxies.
The mean value is calculated as an error-inverse-weighted mean.
Uncertainty of the mean value is estimated from the propagation of errors, taking account of the correlation between contours.
How the errors of isophote shape parameters of each contour are derived is described in the next subsection.
We constrain the semi-major axis range in which the mean value of radial profile is calculated between $2 r_{\mathrm{PSF}}$ and $2 a_{\mathrm{h}}$.
Here, $a_{\mathrm{h}}$ is the half-light semi-major axis calculated from \verb"FLUX_RADIUS" and \verb"ELONGATION" obtained by \verb"SExtractor".
We have confirmed that taking 1.0 or 1.5 $a_{\mathrm{h}}$ instead does not change our result.
Hereafter, isophote shape parameter (e.g., $a_{4}$) simply indicates the mean value.

\subsection{Estimation of Errors of Isophote Shape Parameters}
\label{subsec: est_error}
We estimate statistical errors of isophote shape parameters (e.g., $a_{4}$) arising from random noise of a flux in each pixel,
including photon noise from objects, and background noise such as photon noise due to sky flux and readout noise.
We first estimate the random noise in each pixel, 
then resolve the propagation of noise onto the position of the contour sampling points,
and finally, errors of the isophote shape parameters are estimated by Monte-Carlo simulation.

In the error estimation, the photon noise from objects is estimated from count per pixel by Poisson statistics.
The background noise $\sigma_{\mathrm{bkg}}$ is estimated from background fluctuation per pixel
as we utilize calibrated, sky-subtracted images.
The background noise is estimated from a cut-out image of each target.
All objects detected by \verb"SExtractor" (see Subsection \ref{Subsubsec: Hiz QGs})
are masked before measuring the pixel to pixel fluctuation.
For {\it HST} images where background noise per pixel is correlated,
$\sigma_{\mathrm{bkg}}$ is corrected for the correlated noise
using a simple equation described in Section 2.3 in \citep{Gonzaga+12}.

Once noise per pixel is calculated for each pixel, the positional error, $(\sigma_{x_{n, i}}, \sigma_{y_{n, i}})$,
or error of radial position $\sigma_{r_{n, i}}$,
of a sampling point of a contour can be computed as described in Subsection \ref{subsec: Contour Determination}.
Finally, one sigma uncertainty is estimated by Monte-Carlo simulation using the position errors.
We resample the sampling points of each contour
adding Gaussian random noise to the fitted deviated ellipse with the standard deviation of $\sigma_{r_{n, i}}$.
We resample the sampling points 100 times and repeat fitting deviated ellipse to derive rms scatter of each isophote shape parameter
which we defined as one sigma uncertainty.

\subsection{The Effect of PSF}
\label{subsec: eff_im_qual}
We evaluate the systematic effect of PSF on the measurement of $a_{4}$ in a simulation.
As mentioned in \citet{Pasquali+06}, measured isophote shapes are affected by PSF
for galaxies with small appearent sizes compared to PSF.
We generated artificial images of low- and high-redshift galaxies, convolved them with typical PSF of low- and high-redshift samples,
and measured the $a_{4}$ parameter in the convolved images.
We modeled the galaxies by S\'{e}rsic profiles
with various combinations of the S\'{e}rsic index, axis ratio, and $a_{4}$ parameter: $n=1.0, 2.5, 4.0, 6.5$; $q=0.2, 0.4, 0.6, 0.8$;
$100 \times a_{4}/a=-4.0, -2.0, 0.0, +2.0, +4.0$.
We also changed the galaxy size.
For low-redshift galaxy models, the effective radii are set to $\log(r_{\mathrm{e}}/\mathrm{arcsec})=0.0, 0.5, 1.0, 1.5, 2.0$.
For high-redshift galaxy models, they are set to $\log(r_{\mathrm{e}}/\mathrm{arcsec})=-1.0, -0.5, 0.0, 0.5$.
We note that $1\mathrm{arcsec}$ corresponds to $0.62$kpc at $z=0.031$ and $8.3$kpc at $z=1.2$,
and the ranges of the effective radius well covers the typical sizes of actual galaxies.

\begin{figure*}[htbp]
	\centering
	\hspace{-1.0cm}	
	\includegraphics[width=12.0cm]{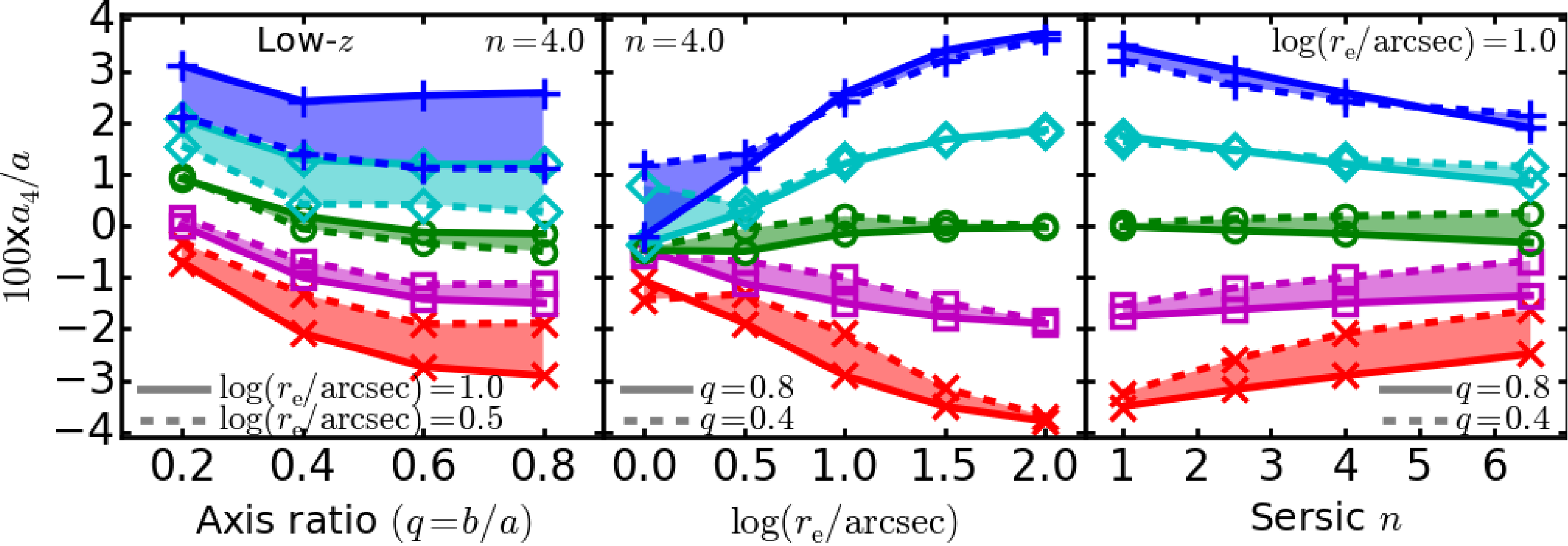}
	\\
	\centering
	\hspace{-1.0cm}	
	\includegraphics[width=12.0cm]{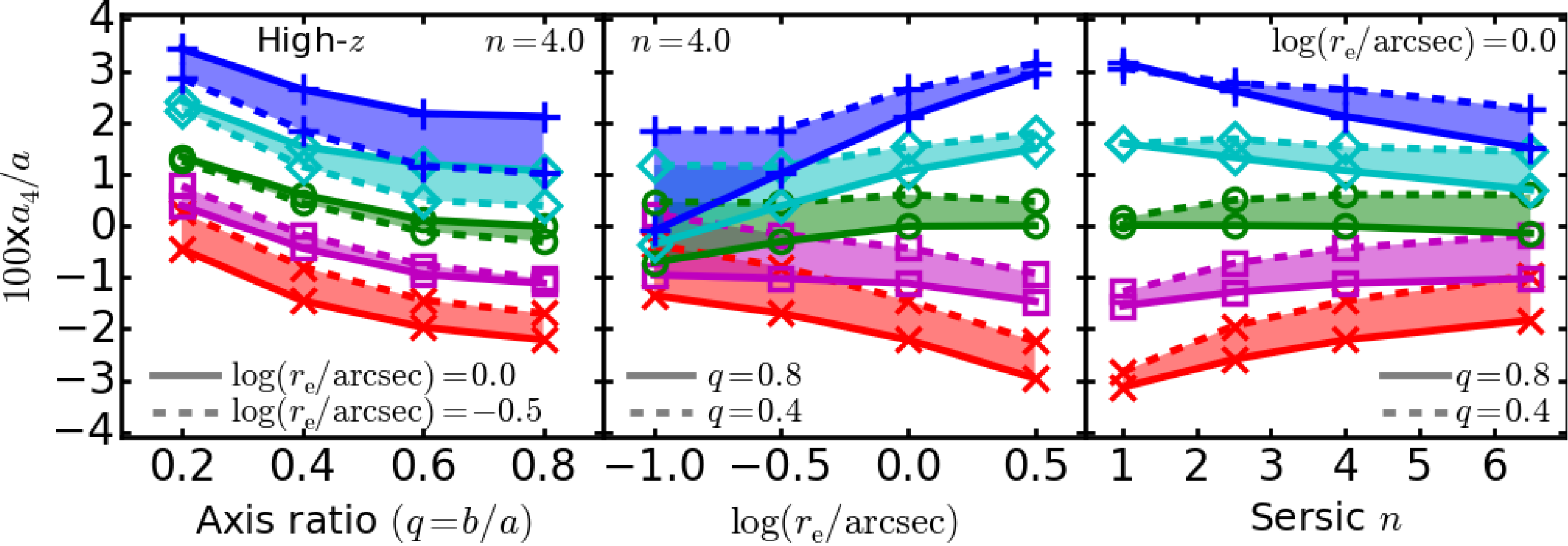}
	\caption{
		PSF-convolved $a_{4}$ parameter of the model galaxies as a function of the input (PSF-deconvolved)
		parameters, the axis ratio ($q$, {\it left}), effective radius ($r_{\mathrm{e}}$, {\it center}), and S\'{e}rsic index ($n$, {\it right}),
		which describes systematic variance of $a_{4}$ due to PSF.
		{\it Upper panels} show the models for low-redshift galaxies whereas {\it lower panels} show those for high-redshift galaxies.
		The shaded region in each panel indicated different input $a_{4}$ parameter.
		The input $a_{4}$ values are, $100 \times a_{4}/a=$-4, -2, 0, +2, and +4,
		shown by {\it red `$\times$' crosses}, {\it magenta squares}, {\it green circles}, {\it cyan diamonds},
		and {\it blue `+' crosses}, respectively.
		\label{fig: PSF_Loz_Hiz}
	}
\end{figure*}

A part of the result is shown in Figure \ref{fig: PSF_Loz_Hiz}
where (PSF-convolved) $a_{4}$ parameters are plotted against the axis ratio, effective radius, and S\'{e}rsic index.
In the left and central panels, only models with $n=4.0$ are shown whereas, in the right panels,
only those with $\log(r_{\mathrm{e}}/\mathrm{arcsec})=1.0$ (0.0) are shown for low-redshift (high-redshift) cases, respectively.
Also, in the left panels, models with $\log(r_{\mathrm{e}}/\mathrm{arcsec})=1.0, 0.5$ (0.0, -0.5) are shown for low-redshift (high-redshift) cases.
In the central and left panels, models with $q=0.8, 0.4$ are shown.

Apparently, the measured (PSF-convolved) $a_{4}$ parameters are affected not only by the axis ratio but also by galaxy size and S\'{e}rsic index.
Moreover, the difference of the measured and intrinsic values of $a_{4}$ depends on the intrinsic $a_{4}$ value.
The dependence of PSF effect on these parameters can be summarized as follows.
First, the isophote shapes tend to be measured as disky especially for flattened galaxies with $q\lesssim0.5$.
Second, galaxies with the large apparent size are not suffered from the PSF effect,
but for small galaxies, the measured isophote shape tends to be round as the relative PSF size to the galaxy become larger.
Third, for galaxies with larger S\'{e}rsic index, the isophote shape tends to be rounder,
because the flux is more concentrated and affected by PSF more strongly.
Finally, the difference of the measured and intrinsic $a_{4}$ value becomes larger with increasing absolute value of $a_{4}$.
\citet{Pasquali+06} show the dependance of the PSF effect on the axis ratio for $a_{4}=0$, and our simulation gives the consistent result.
We note that only small axis ratios cause systematic effect for disky and boxy classification,
i.e., boxy intrinsic shape affected by the PSF may be measured as disky.
Other parameters, small sizes and large S\'{e}rsic indices, makes absolute $a_{4}$ value small, but do not affect the classification.

The effect of PSF on the isophote shape measurement depends on many parameters and complex,
and it is rather difficult to correct the $a_{4}$ value for the effect.
However, as we restricted the redshift range of the low-redshift sample so that the PSF size become consistent between low and high redshifts,
and the sizes, axis ratios, S\'{e}rsic indices do no change between the high- and low-redshift ETG samples very much,
the PSF effect is probably similar in both low- and high-redshift samples.
Thus, we decided not to correct the measured isophote shape parameters in this study.


\section{Results}
\label{Sec: Results}

\begin{figure}[htbp]
	\plotone{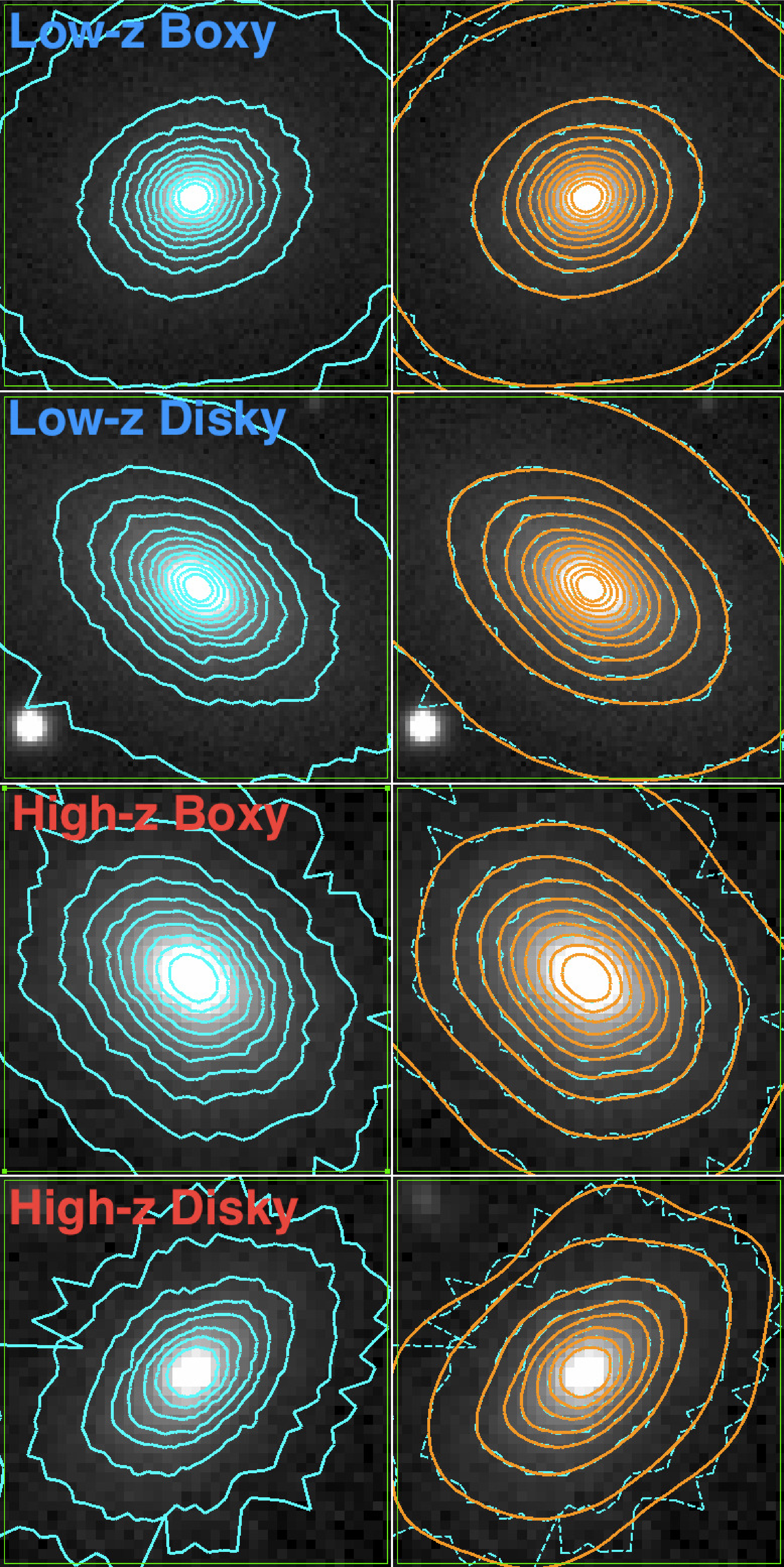}
	\caption{
		Examples of low-redshift boxy and disky ETGs,
		and high-redshift boxy and disky ETGs are shown from top to bottom.
		In the {\it left panels}, the isophote contours are overlaid with {\it cyan solid lines}.
		In the {\it right panels}, the fitted deviated ellipses are overlaid with {\it orange solid lines}
		whereas the isophote contours are shown in {\it cyan dashed lines}.
		The size of the image is $30\times 30$ and $2\times2\mathrm{arcsec^2}$ for low and high redshift, respectively.
		\label{fig: Profimg_Loz_Hiz}
	}
\end{figure}

In this section, we present the results of the isophote shape measurements
for the low- and high-redshift samples.
In Figure \ref{fig: Profimg_Loz_Hiz},
examples of the isophote shape measurement of the low- and high-redshift ETGs are shown.
The isophote contours are well fitted by the deviated ellipses.

\subsection{Relation between $a_{4}$, Mass, and Size}
\label{subsec: a4_Ms_Re}
\begin{figure*}[htbp]
	\vspace{0\baselineskip}
	\centering
	\hspace{-1.0cm}	
	\includegraphics[width=11.0cm]{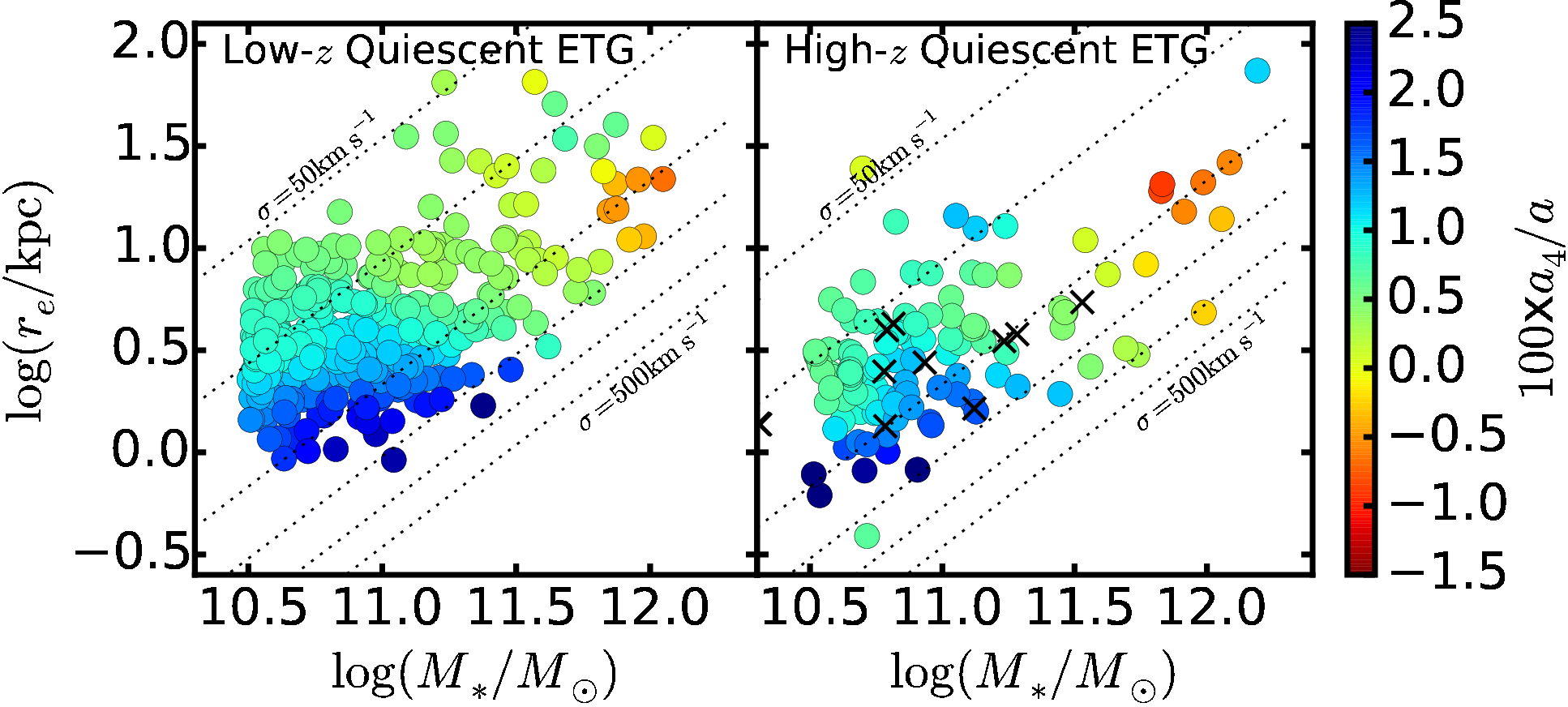}
	\caption{
		Distribution of the low- ({\it left}) and high-redshift ({\it right}) quiescent ETG samples on the mass-size ($M_{*}$-$r_{e}$) plane.
		Color code indicates the $a_{4}$ parameter which is locally averaged around the data point on the plane.
		$r_{e}$ is the effective radius measured with GALFIT.
		Black crosses in right panel indicates ETGs at $z\sim0.5-1.1$ from \citet{Pasquali+06}.
		Dotted lines indicate lines of constant velocity dispersion, $\sigma=50, 100, 200, 300, 400, 500 \mathrm{km\ s^{-1}}$ from left to right,
		assuming the virial relation $M_{*}=5.0 \times \sigma^2 r_{e} / G$.
		\label{fig: MP_a4}
	}
	\vspace{0\baselineskip}
\end{figure*}

In Figure \ref{fig: MP_a4}, we show the distribution of the low- and high-redshift ETG samples on the mass-size ($M_{*}$-$r_{e}$) plane
with the $a_{4}$ parameter color coded.
In this figure, $a_{4}$ values are locally averaged around each data point within the plane to see global trends.
For both the low- and high-redshift samples, we confirm a well known correlation between the $a_{4}$ parameter and mass, i.e., ETGs tend to be boxy with increasing stellar mass.
Both in low- and high-redshift samples, the main population of ETGs changes from disky to boxy
at a critical stellar mass of $\log(M_{*}/M_{\odot}) \sim 11.5$, and the massive end is dominated by boxy ETGs.
Although the stellar mass of the high-redshift galaxies has a large uncertainty as we can use only two-band photometry,
the critical mass at which main population of ETG changes from disky to boxy is in good agreement with
the characteristic mass of $\log(M_{*}/M_{\odot}) \sim 11.3-11.5$ at which the dynamical property of nearby ETGs changes from fast to slow rotators \citep{Emsellem+11, Cappellari+13b}.

Figure \ref{fig: MP_a4} also shows a global trend for low mass galaxies ($\log(M_{*}/M_{\odot})\la11.5$) that
they generally become more disky with decreasing size compared at the same mass (i.e., with increasing velocity dispersion) for low- and high-redshift samples.
For high mass galaxies ($\log(M_{*}/M_{\odot}) \ga 11.5$), 
as we do not have galaxies with small $r_{e}$,
we can not examine the size dependance of the $a_{4}$ parameter.
For the most massive galaxies  with $\log(M_{*}/M_{\odot}) \ga 11.7$, disky galaxies tend to be larger ($\log(r_{e}/\mathrm{kpc})\gtrsim1.5$) in the low-redshift sample.
We have one galaxy for the high-redshift sample in this stellar mass and effective radius range and it is disky.
Our high-redshift sample lacks massive ($\log(M_{*}/M_{\odot}) \ga 11.5$), large ($\log(r_{e}/\mathrm{kpc})\gtrsim1.5$) ETGs
(in other words, massive but rather small velocity dispersion of $\sim100\ \mathrm{km \cdot s^{-1}}$ which is indicative of a disk dominated system).
This may be the result of building up of quiescent disk dominated galaxies quenched in $z<1$ from blue population at $z\sim1$.
The situation does not change if we plot this figure with quiescent galaxies before the morphological ETG selection.
We have checked the image of massive disky galaxies in our low- and high-redshift ETG samples
and found that three out of fifteen low-redshift galaxies have spiral like feature but none of seven high-redshift ones has such feature.
The incompleteness of spectroscopy for high-redshift sample is less likely to be the reason for the lack of massive small dispersion galaxies
as morphological selection is not applied for the selection of spectroscopic targets.

Figure \ref{fig: MP_a4} well illustrates the transition of ishophote shapes of ETGs
within the mass-size plane as a function of constant velocity dispersion,
which is similar to the transition of dynamical properties described in Figure 8 in \citet{Cappellari+13b}.
We also plot $M_{*}$ and $r_{e}$ of ETGs in HST ACS Ultra Deep Field at $z\sim0.5-1.1$ whose isophote shapes are studied by \citet{Pasquali+06}.
We estimate the stellar mass from the four-band photometry ($g$, $r$, $i$, $z$) presented in Table 1 in \citet{Pasquali+06} using BC03 SSP fit with the Salpeter IMF as for our low-redshift sample,
and effective radii is taken from \citet{van_der_Wel+14}.
Our high-redshift quiescent ETG sample and those of \citet{Pasquali+06} occupy similar loci on the mass-size plane,
although our sample has very large and massive ellipticals as we selected ETGs in massive galaxy clusters.

\begin{figure*}[htbp]
	\vspace{0\baselineskip}
	\centering
	\hspace{-1.0cm}	
	\includegraphics[width=11.0cm]{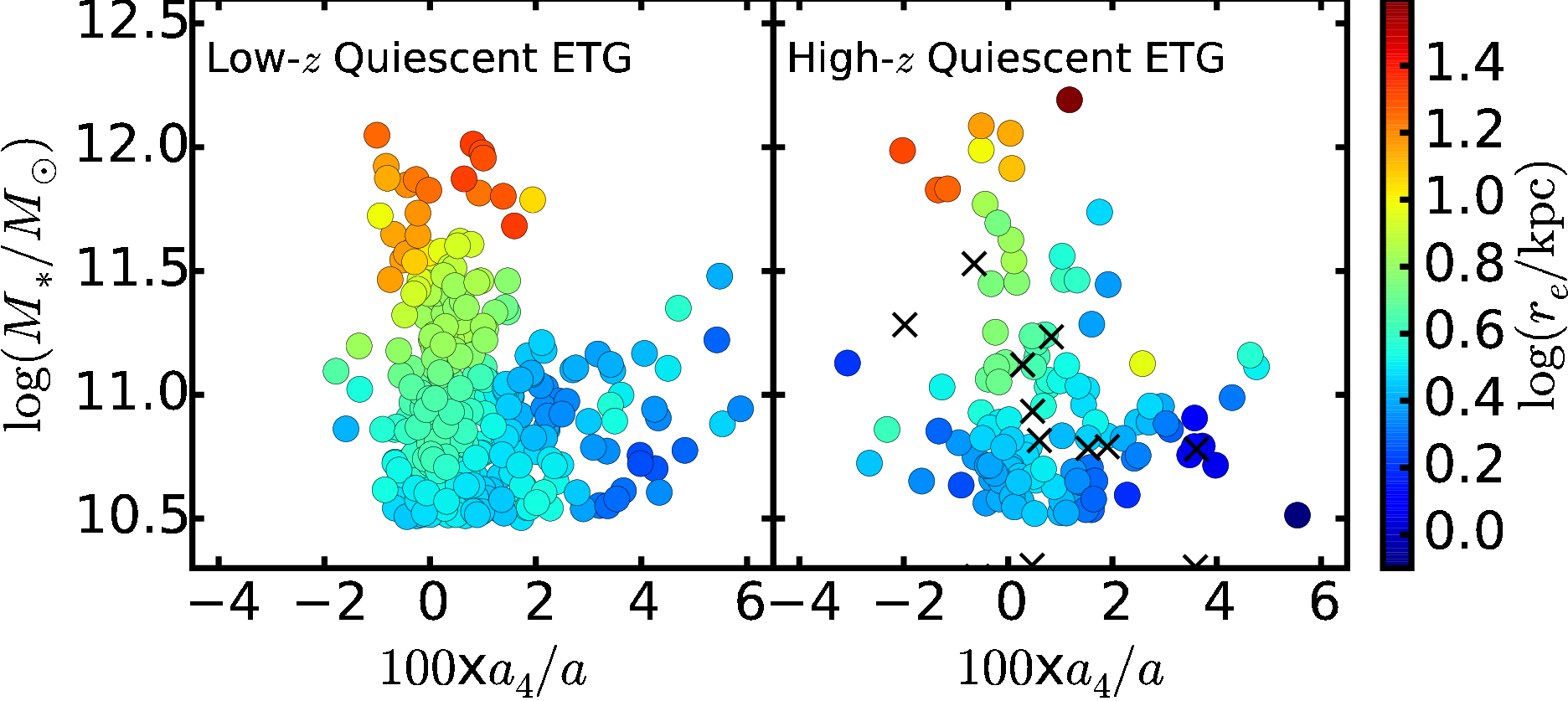}
	\caption{
		Distribution of the low- ({\it left}) and high-redshift ({\it right}) quiescent ETGs on $M_{*}$-$a_{4}$ plane.
		Color code indicates the locally averaged effective radius.
		Black crosses in right panel indicates ETGs at $z\sim0.5-1.1$ from \citet{Pasquali+06} as in Fig. \ref{fig: MP_a4}.
		\label{fig: a4_Ms}
	}
	\vspace{0\baselineskip}
\end{figure*}

In Figure \ref{fig: a4_Ms}, we exchange the axes of Figure \ref{fig: MP_a4}, showing the distribution of the ETG samples on the $M_{*}$-$a_{4}$ plane
with the locally averaged effective radius $r_{e}$ color coded.
Disky ETGs are most frequent around $\log(M_{*}/M_{\odot}) \sim 10.5$, but are rare in $\log(M_{*}/M_{\odot}) > 11.5$ in both low- and high-redshift samples.
On the other hand, boxy ETGs appear in all stellar mass ranges.
Hence, the transition of main population from disky to boxy with increasing stellar mass is due to disappearance of disky galaxies in the massive end.
We also plot $a_{4}$ and $M_{*}$ of the $z\sim0.5-1.1$ ETGs from \citet{Pasquali+06}.
The $a_{4}$ parameters are taken from Table 2 in \citet{Pasquali+06} where the isophote shape parameter is not corrected for PSF as with ours.
Our high-redshift ETG sample and \citet{Pasquali+06} sample have similar $M_{*}$-$a4$ distribution.

\begin{figure*}[htbp]
	\vspace{0\baselineskip}
	\centering
	\hspace{-1.0cm}	
	\includegraphics[width=11.0cm]{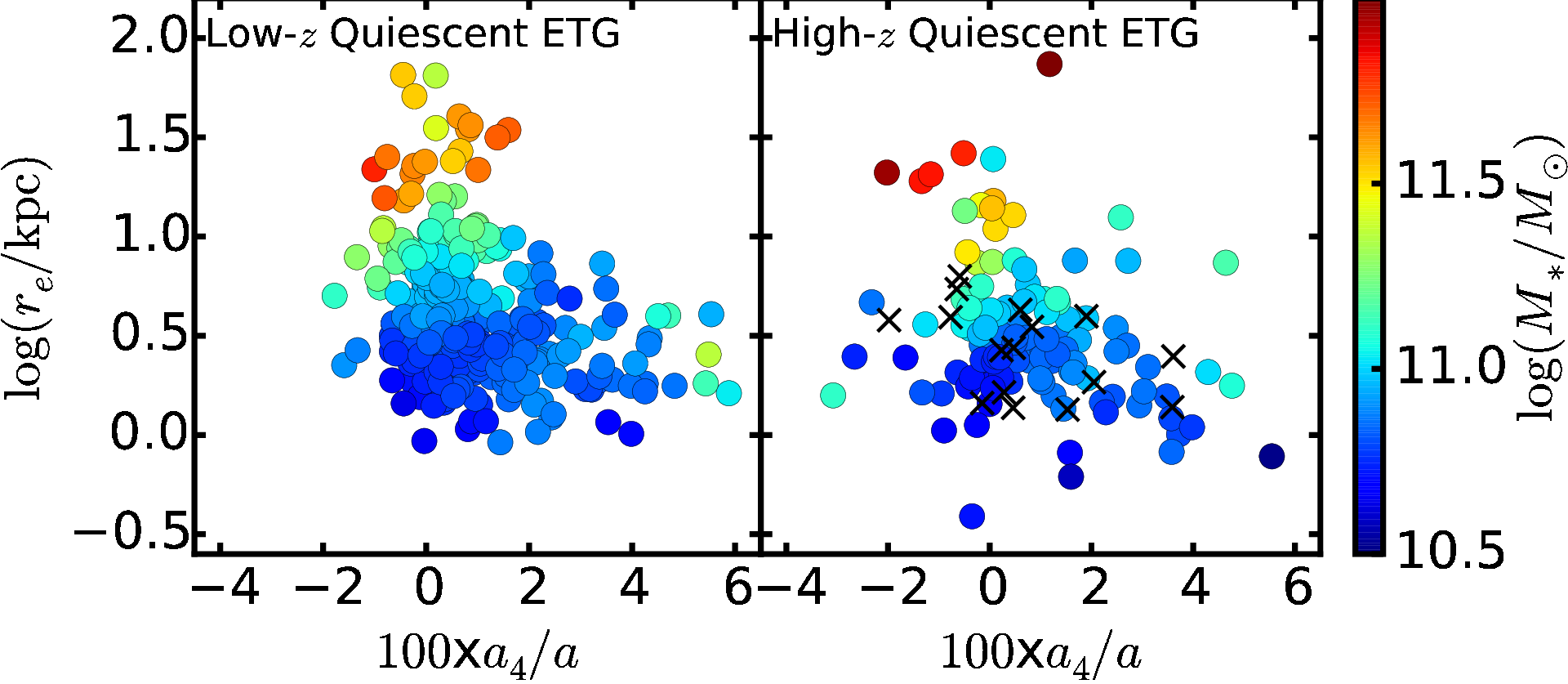}
	\caption{
		Distribution of the low- ({\it left}) and high-redshift ({\it right}) quiescent ETGs on $r_{e}$-$a_{4}$ plane.
		Color code indicates the locally averaged stellar mass.
		Black crosses in right panel indicates ETGs at $z\sim0.5-1.1$ from \citet{Pasquali+06} as in Fig. \ref{fig: MP_a4}.
		\label{fig: a4_Rh}
	}
	\vspace{0\baselineskip}
\end{figure*}

In Figure \ref{fig: a4_Rh}, we show the distribution of the ETG samples on the $r_{e}$-$a_{4}$ plane where the color code indicates
the locally averaged stellar mass.
As expected from Figures \ref{fig: MP_a4} and \ref{fig: a4_Ms},
the majority of ETG with $\log(r_{e}/\mathrm{kpc)} \lesssim 0.5$ is disky whereas in $\log(r_{e}/\mathrm{kpc)} \gtrsim 0.5$,
boxy ETGs become frequent in both low- and high-redshift samples.
As there is the size-mass relation, this critical size correspond to $\log(M_{*}/M_{\odot}) \sim 11.5$ (see Figure \ref{fig: MP_a4}).


\subsection{Disky ETG Fraction}
\begin{figure}[htbp]
	\vspace{0\baselineskip}
	\plotone{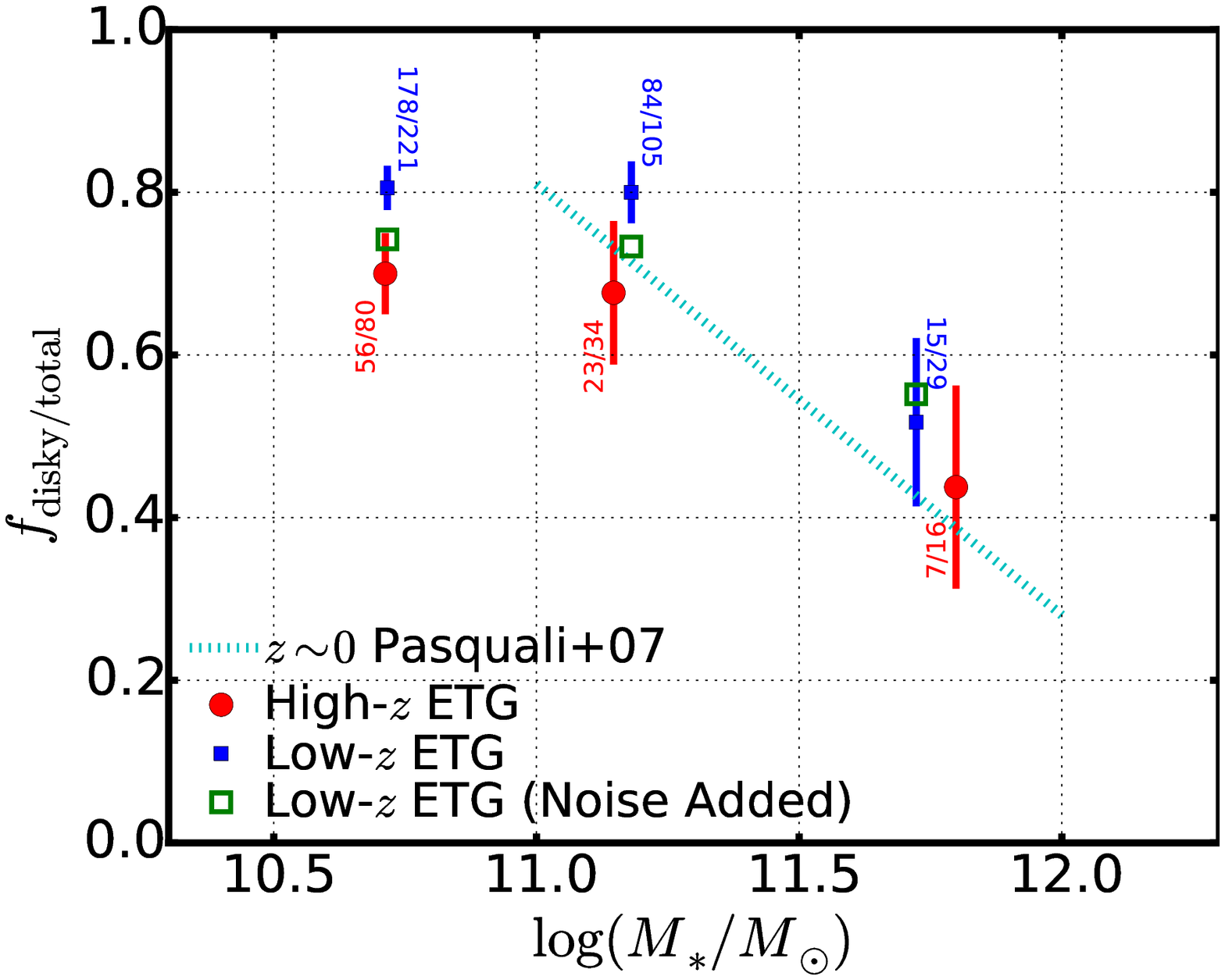}
	\caption{
		Disky-to-total fraction for low- ({\it{blue squares}}) and high-redshift ({\it red circles})  quiescent ETGs as a function of stellar mass.
		{\it Green open squares} represent the expected disky-to-total fraction of the low-redshift ETGs when the uncertainty of the $a_{4}$ parameter
		is comparable to that of the high-redshift ETGs (see text and Appendix).
		In each mass bin, the position in $x$ axis shows the median stellar mass.
		The numbers on each point indicate the number of disky and total ETGs in each mass bin.
		{\it Cyan dotted line} indicates the disky-to-total fraction of $z\sim0$ ETGs obtained by \citet{Pasquali+07}.
		\label{fig: a4F_d2t}
	}
\end{figure}

In Figure \ref{fig: a4F_d2t}, we show the fraction of disky ETGs as a function of stellar mass.
We classified ETGs with $a_{4}>0$ as disky.
The fraction is calculated in three mass bins, $10.5 \leq \log(M_{*}/M_{\odot}) < 11.0$,  $11.0 \leq \log(M_{*}/M_{\odot}) < 11.5$, and  $11.5 \leq \log(M_{*}/M_{\odot})$.
The error bars indicate 16 and 84 percentile computed by 1000-time bootstrap resampling where the same number of galaxy is randomly resampled and the disky fraction is computed for each resampling.
For comparison, we also plot the linear function fit to disky-to-total fraction of nearby ETGs from Equation (6) in \citet{Pasquali+07}, applying our cosmology parameter.
The fraction of our low- and high-redshift ETGs is consistent with the linear function.
Note that the stellar mass of ETGs in \citet{Pasquali+07} spans from $\log(M_{*}/M_{\odot}) \sim 11.0$ to $\log(M_{*}/M_{\odot}) \sim 11.7$.

We do not find significant differences in the disky-to-total fraction.
For the highest mass bin ($11.5 \leq \log(M_{*}/M_{\odot})$), the disky fraction is consistent between low- and high-redshift samples within $\sim0.09$.
The fraction appears higher in low-redshift but the difference is insignificant.
If we exclude the three spiral galaxies in the low-redshift sample as contamination (see Subsection \ref{subsec: a4_Ms_Re}),
the disky fraction of the low-redshift sample becomes 0.41 which is closer to that of the high-redshift (0.43) than original low-redshift value of 0.52.
Therefore, we conclude that disky fraction for massive ETGs probably stays the same at $z\sim1$ and 0 in cluster environment. 
Since the uncertainty of the disky fraction for massive ETGs arise mainly from the small number of samples
(the measurement error of $a_{4}$ is not large in this mass range),
we need larger sample to confirm the conclusion with higher accuracy.

For the lower mass bin ($10.5 \leq \log(M_{*}/M_{\odot}) < 11.0$ and $11.0 \leq \log(M_{*}/M_{\odot}) < 11.5$ ), one might find that the disky fraction of the high-redshift sample is significantly lower than that of low-redshift sample.
However, the difference may be the result of the large measurement error in $a_{4}$ of the high-redshift sample
which modifies the shape of the distribution of $a_{4}$ and can bias the disky fraction into smaller value.
This kind of bias is known as Eddinton bias \citep{Eddington13, Teerikorpi04}.
In Appendix, we present the detail of the effect of measurement error of $a_{4}$ on the disky fraction.
If we bring the measurement error of $a_{4}$ of the low-redshift sample equivalent to that of the high-redshift by adding gaussian noise to $a_{4}$ distribution of the low-redshift sample ({\it green open squares} in Figure \ref{fig: a4F_d2t}),
the disky fraction of low- and high-redshift sample would become consistent within $\sim0.05$ (see Appendix).
In order to detect possible differences of the disky fraction,
we need to increase the sample size of both the low- and high-redshift samples to $\sim$300, 400, and 800 for the highest, intermediate, and lowest mass bins,
or, in order to compare disky fraction without the Eddington bias,
we need to reduce the measurement error of $a_{4}$ of high-redshift galaxies down to equivalent amount of that of low-redshift galaxies
Thus, we need much deeper imaging of high-redshift galaxies with spatial resolution of space-based telescope, but our imaging data are one of the deepest ones with {\it HST}.
In the near future, {\it James Webb Space Telescope} will enable us to study isophote shapes for lower mass high-redshift galaxies with much higher accuracy.

\subsection{The uncertainty of the $a_{4}$ parameter}
\begin{figure}[htbp]
	\vspace{0\baselineskip}
	\plotone{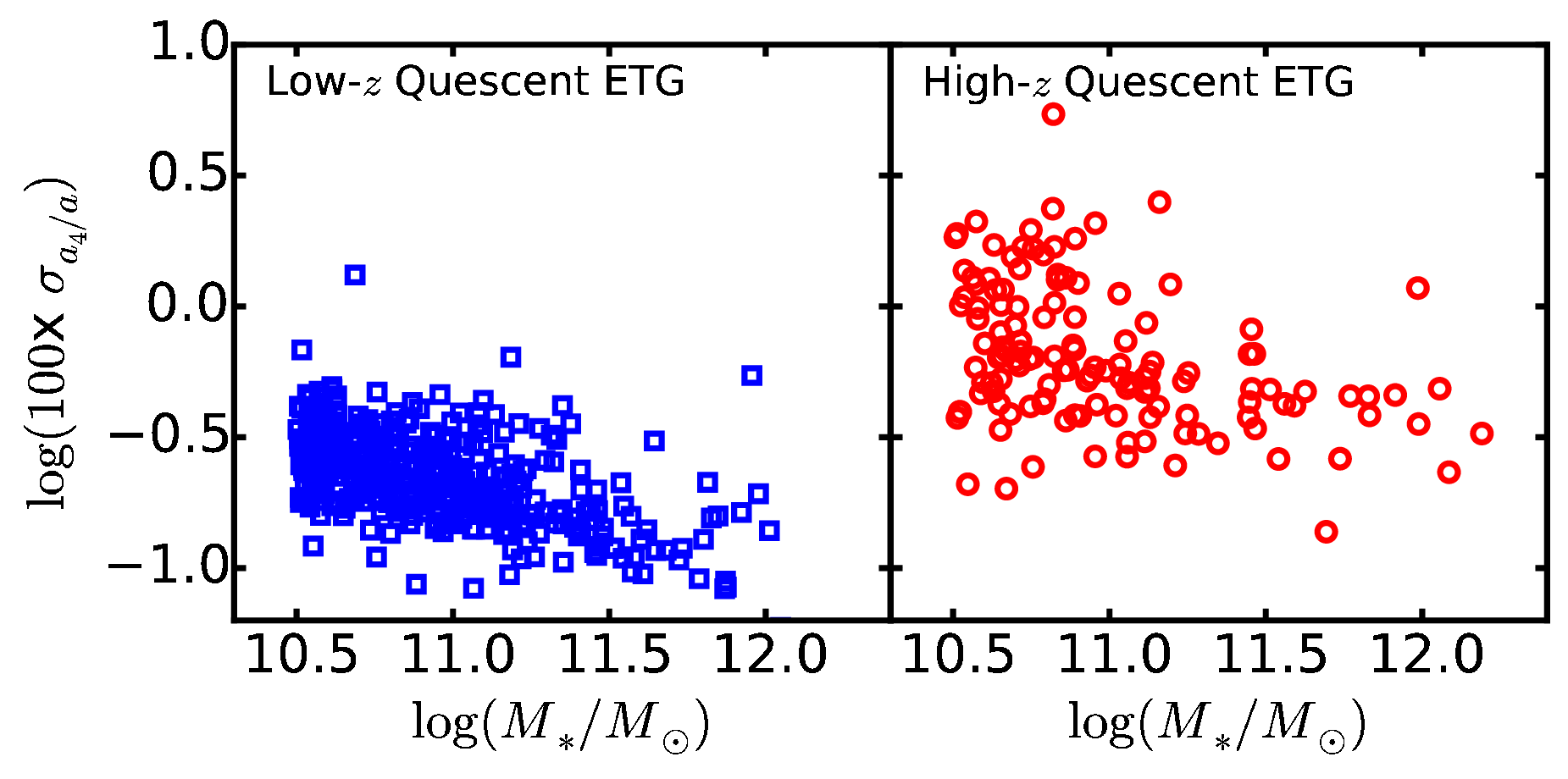}
	\caption{
		Measurement error of $a_{4}$ as a function of the stellar mass
		for the low- ({\it left}) and high-redshift ({\it right}) quiescent ETGs.
		\label{fig: a4sig_Ms}
	}
	\vspace{0\baselineskip}
\end{figure}

We show the measurement error of the $a_{4}$ parameter as a function of stellar mass in Figure \ref{fig: a4sig_Ms}.
We can see that the error tends to decrease with increasing stellar mass .
For the high-redshift quiescent ETGs, the average measurement uncertainties are 
$100 \times \sigma_{a_{4}/a} \sim 0.7, 0.5, 0.4$ at $\log(M_{*}/M_{\odot}) \sim 10.5, 11.5, 12.0$, respectively.
For the low-redshift ones, the uncertainties are $100 \times \sigma_{a_{4}/a} \sim 0.3, 0.2, 0.1$ at $\log(M_{*}/M_{\odot}) \sim 10.5, 11.5, 12.0$.

\subsection{The relation between the $a_{4}$ parameter and axis ratio}
\begin{figure}[htbp]
	\vspace{0\baselineskip}
	\plotone{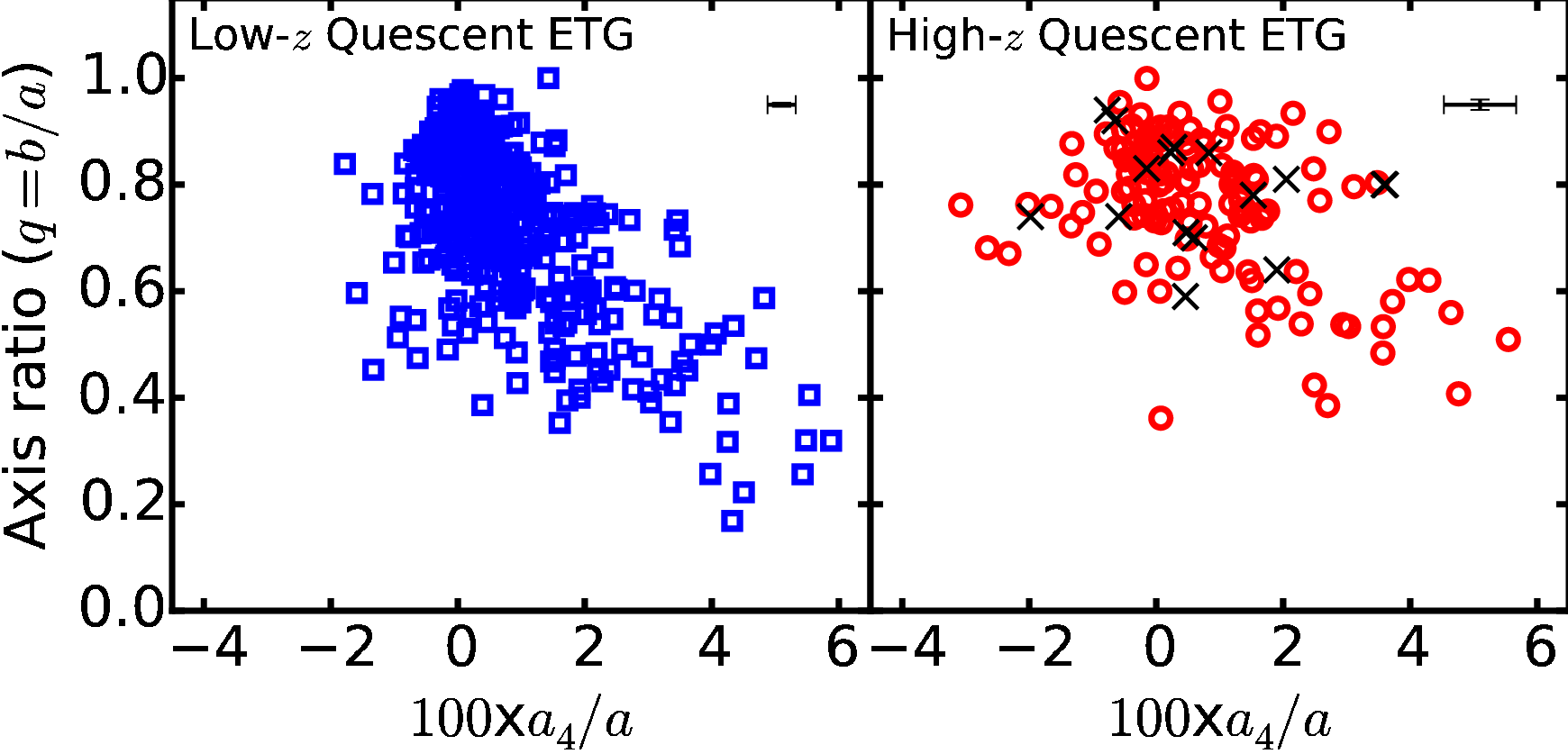}
	\caption{
		Axis ratio $q=b/a$ plotted against the isophote shape parameter $a_{4}$
		for the low- ({\it left}) and high-redshift ({\it right}) quiescent ETGs.
		{\it Black crosses} in the right panel indicates ETGs at $z\sim0.5-1.1$ from \citet{Pasquali+06}.
		Here, axis ratios are measured in the same way as $a_{4}$ and not deconvolved with PSF.
		\label{fig: a4_CERPq}
	}
\end{figure}

In Figure \ref{fig: a4_CERPq}, we plot the axis ratio against the isophote shape parameter $a_{4}$.
The absolute value of the $a_{4}$ parameter tends to be small when the axis ratio is close to unity,
and ETGs with small axis ratios tend to be disky for both the low- and high-redshift samples.
These trends are already known for nearby ETGs \citep{Bender+89}.
We also plot the field ETGs at $z\sim0.5-1.1$ from \citet{Pasquali+06}.
Our high-redshift ETGs cover the similar region as theirs.

%

\section{Discussion}
\label{Sec: Discuss}
We discuss the evolution of ETGs in massive clusters between $z\sim1$ and 0,
based on the isophote shapes as well as other galaxy properties.
A theoretical study predicts the dynamical evolution of ETGs \citep{Khochfar+11},
and observational studies have found size evolution and morphological evolution in terms of the axis ratios and S\'{e}rsic indices \citep[e.g.,][]{Bundy+10, Damjanov+11, Newman+12, Cimatti+12, YYChang+13a, YYChang+13b, van_der_Wel+14, Delaye+14}.
However, we do not find significant evolution in disky-to-boxy fraction between $z\sim1$ and $\sim0$,
which suggests the isophote shapes and probably the dynamical properties of ETGs are already in place at $z>1$.
%
We would like to investigate the implication from these observational facts in this section.

\subsection{The Effect of PSF on the Disky Fraction}
First, we discuss the effect of the PSF on the disky to total fraction to make sure that the PSF does not affect the discussion about the evolution of the disky ETG fraction.
As we have mentioned in Section \ref{subsec: eff_im_qual},
the measurement of the $a_{4}$ parameter is affected by the PSF.
So, the difference of the PSF between low- and high-redshift samples may introduce some systematics into the evolution of the disky fraction.
As the PSF sizes of low- and high-redshift samples are comparable in physical scales,
what may matter are different sizes, S\'{e}rsic indices, and axis ratios of the sample galaxies.

First, the high-redshift galaxies have on average smaller sizes in physical scale than the low-redshift one.
However, smaller sizes only make absolute value of observed $a_{4}$ small, and do not change disky to boxy (i.e., the sign of $a_{4}$)
and vice versa.

Second, the high-redshift galaxies have similar S\'{e}rsic indices to the low-redshift ETGs in our sample.
Although the distribution of the S\'{e}rsic index of the high-redshift ETGs seems to spread in wider range (Figure \ref{fig: M_n_all}),
the S\'{e}rsic index does not alter the sign of the $a_{4}$ parameter.

Finally, the axis ratios of the high-redshift sample is smaller than those of the low-redshift sample
although the statistical significance is not so high in our samples.
This would increase the observed $a_{4}$ parameter greater for the high-redshift sample than for low-redshift,
and the disky fraction of the high-redshift sample may be over-estimated.
If the disky fraction increases at $z\sim1$ as predicted in \citet{Khochfar+11}
and the smaller axis ratio of high-redshift galaxy affect the disky-boxy classification,
we would observe more significant evolution in the disky fraction but we do not.

Therefore, even if we take account of the effect of the PSF, it can not explain the fact that there is no evolution in the disky fraction.
Moreover, as the differences of galaxy sizes, S\'{e}rsic indices, and axis ratios are very small between the low- and high-redshift samples,
the PSF probably affects the isophote shape measurement equally in low- and high-redshift samples,
and does not probably introduce systematics in the evolution of the disky fraction.

\subsection{Isophote Shapes of Massive ETGs}
In the previous section,
we present that massive ETGs with $\log(M_{*}/M_{\odot})\gtrsim11.5$ are basically boxy for both $z\sim1$ and 0 samples.
This result indicates that their boxy isophote shapes and probably dynamical properties are already in place at  $z>1$.
One important point is that the critical mass at which the main population of ETG changes from disky to boxy
is consistent between $z\sim1$ and $0$ with the value of $\log(M_{\rm{crit}}/M_{\odot}) \sim 11.5$.
This is consistent with the idea that the origin of boxy ETGs is a kind of mass quenching \citep[see][and references therein]{Kormendy+09}.
When gas accretes onto massive galaxies, a shock develops, the gas is heated to the virial temperature, and star formation is quenched.
The hot gas is maintained as hot by additional accretion \citep{DekelBirnboim06, DekelBirnboim08} and AGN feedback \citep{Best+06, Best07a, Best07b}.
As a result, any mergers become dry for massive galaxies with $M_{*} \gtrsim M_{\rm{crit}}$, and merger remnants tend to be boxy.
The idea is supported by observational facts that blue star-forming galaxies are less massive than $\sim 10^{11} M_{*}$
in the local universe \citep[e.g.,][]{Baldry+04, Baldry+06} and in the intermediate- to high-redshift universe \citep[e.g.,][]{Bell+04, Faber+07}.
In addition, existence and non existence of X-ray emitting gas in massive and less massive ETGs \citep{Bender+89, Pellegrini99, Pellegrini05, EllisO'Sullivan06}
also make the mass quenching scenario attractive \citep[see][for summary]{Kormendy+09}.

Theoretical studies done by \citet{DekelBirnboim06, DekelBirnboim08} suggest that the critical halo mass is $\log(M_{\rm{halo}}/M_{\odot})$ $\sim 12$
which corresponds to the stellar mass of $\log(M_{\rm{*}}/M_{\odot})$ $\sim11.2$
using a baryon-to-total mass ratio of $1/6$ \citep{Komatsu+09} as presented in \citet{Kormendy+09}.
This mass is in good agreement with the critical mass for boxy-disky transition.
If the mass quenching is the main origin of massive boxy ETGs, the critical mass is basically constant regardless of redshifts.
Some authors suggest these massive ETGs experience a few to several major mergers at $z<1$ \citep{Lidman+12, Lidman+13, Shankar+15}.
These mergers should be dry as wet mergers convert boxy ETGs into disky \citep{Khochfar+05, Naab+06}.
It is important to examine whether the critical mass is constant regardless of redshifts even in higher redshifts, but it is beyond the scope of this paper.

The environment may also play a role for the characterization of the isophote shapes \citep[e.g,][]{Shioya+93, Hao+06} and dynamical properties of ETGs \citep[e.g.,][]{Cappellari+11b}.
\citet{Hao+06} present that boxy ETGs favor denser environment while disky ETGs favor more isolated environment.
IFS studies have revealed that massive, slowly rotating ellipticals are preferentially found in the central region of galaxy clusters \citep{Cappellari+11b, Houghton+13, D'Eugenio+13, Scott+14}.
We investigate the disky fraction for ETGs in the inner and outer region of each cluster,
separating the high- and low-redshift samples by the cluster centric radius ($r_{\rm{CL}}$) with $\log(r_{\rm{CL}}/r_{200}) \leq -0.4$ or $> -0.4$.
As in the previous section, we calculate the disky fraction in three mass bins.
We have confirmed that the disky fraction increases in the outer region in all mass bins for the low-redshift sample.
However, we can not obtain meaningful result for the high-redshift sample due to small number of sample especially in the outer region.
For better understanding about the environmental effect on the disky fraction at high-redshift,
we need to increase the number of sample in the outer region of galaxy clusters or less dense environment.


\subsection{Morphological Evolution of Less Massive ETGs in Dense Environment}
For less massive ETGs with $\log(M_{*}/M_{\odot})\lesssim11.5$ where disky ETG is the dominant population,
the disky fraction is consistent between $z\sim1$ and $0$ in this mass range,
and the higher disky (fast rotator) fraction at high redshift predicted by a theoretical study \citep{Khochfar+11} is not observed.

\citet{YYChang+13a, YYChang+13b} study evolution of the intrinsic shape (oblate or triaxial) of ETGs residing in the field environment using distributions of projected axis ratios.
They present that for ETGs with $10.5<\log(M_{*}/M_{\odot})<11.5$, the fraction of oblate ETGs is higher at $z>1$.
If we interpret the oblate-triaxial classification as disky-boxy (i.e., fast-slow rotator) classification,
the fact that no significant increase in the disky fraction is found in this study seems to be inconsistent with the increase of the oblate fraction.
However, as our ETGs reside in very massive clusters, we need to take account of the evolution of ETGs in such an environment.
For example, evolution of the size of ETGs is different between field and cluster environments.
Field ETGs at $z\sim1$ smaller sizes than the local counter parts \citep[e.g.,][]{Trujillo+07, Damjanov+11, Cimatti+12, Newman+12, van_der_Wel+14}.
However, \citet{Delaye+14} report that the average size of ETGs living in massive clusters at $0.8 < z < 1.5$ appear to be on average 1.5 times larger than those residing in the field at similar redshifts when compared at the same mass.
The distribution of the axis ratio may also be different between environments.
Although axis ratios of our high-redshift ETGs are smaller than those of our low-redshift ones,
the axis ratio distribution of the high-redshift sample is not as flat as the field sample at $z\sim1$ in \citet{YYChang+13a, YYChang+13b}.
While the axis ratio histogram of the field sample at $1<z<2.5$ in the stellar mass range of $10.8<\log(M_{*}/M){\odot})<11.5$ has a peak around $q\sim0.6$ \citep[see Figure 5 in][]{YYChang+13b}, 
our cluster ETGs at $z\sim1.2$ has a peak at $q\sim0.7-0.8$.
Thus, the cluster ETG sample probably have a lower oblate-to-triaxial fraction than the field sample.
The difference of the sizes and axis ratios between clusters and fields indicates that the morphological evolution of ETGs should be weaker in a dense environment at $z<1$.
It may be worthwhile to investigate evolution of isophote shapes of filed ETGs in future work as their morphological evolution is expected to be more pronounced.
In the rest of this section, we focus on the evolution of ETGs in a dense environment.

Although the size evolution of ETGs seems to be weak at $z<1$,
some authors report the evolution of morphology of ETGs in massive clusters between $z\sim1$ and 0.
\citet{Mei+12} present that galaxies in the Lynx super cluster at $z=1.3$ show high fractions of red, bulge-dominated disk galaxies.
\citet{Cerulo+14} claim that the main population of intermediate mass ETGs in massive cluster changes from bulge-dominated disk galaxies at $z\sim1$ to elliptical galaxies at $z\sim0$.
\citet{De_Propris+15} also report the evolution of the axis ratio and the S\'{e}rsic index but no size evolution,
comparing ETGs in massive clusters at $z\sim1$ and those in Virgo cluster.
They conclude that ETGs in dense environment at $\sim1$ have similar size but are on average more flatten and less concentrated than local ones.
\citet{Bundy+10} show an increase of massive ($M_{*} \gtrsim10^{11}M_{\odot}$) ETGs
and a decline of $\log(M_{*}/M_{\odot}) \sim 11$ bulge-dominated spiral galaxies with Sersic index $1.25<n<2.5$ from $z\sim1$ to $0$
although their sample is taken from the COSMOS field.
They infer that at least 60\% of the bulge-dominated spiral galaxies are transformed into ETGs (e.g., S0s) on the scale of 1-3 Gyr,
and that these transformations might occur as a single major merger event or through multiple evolutionary stages,
including disk disruption by minor mergers or accretion of cold gas in star-forming galaxies.

\citet{Feldmann+11} simulated the formation of a group of galaxies,
and found that while elliptical galaxies are formed in mergers at $z>1$, before the merging progenitors fall within the virial radius of the group,
unmergerd disk galaxies are turned into red-and-dead disks in the group environment due to shutting down of gas accretion and stripping of gas.
Thus, based on their simulation, the disk-like ETGs in massive clusters at $z\sim1$ may originate from quenched disk galaxies.
\citet{Carollo+13, Carollo+14aph} present that quenching and fading of disk galaxies may be responsible for the apparent size evolution of ETGs as a function of time.

Possible mechanisms for fading of disk galaxies into ETGs in a dense environment is not only mergers but also secular evolution.
\citet{Oesch+10} found that intermediate mass ($\log(M_{*}/M_{\odot})\lesssim11$) disk galaxies are present at $z=1$ in the COSMOS, but vanish by $z=0.2$,
and suggest that these disk galaxies may be transformed into ETGs by secular evolution
as the merger rate is too low to account for the observed decrease in their space densities.

\begin{figure*}[htbp]
	\vspace{0\baselineskip}
	\centering
	\hspace{-0.5cm}
	\centering 
 	\leavevmode 
 	\includegraphics[width=8cm]{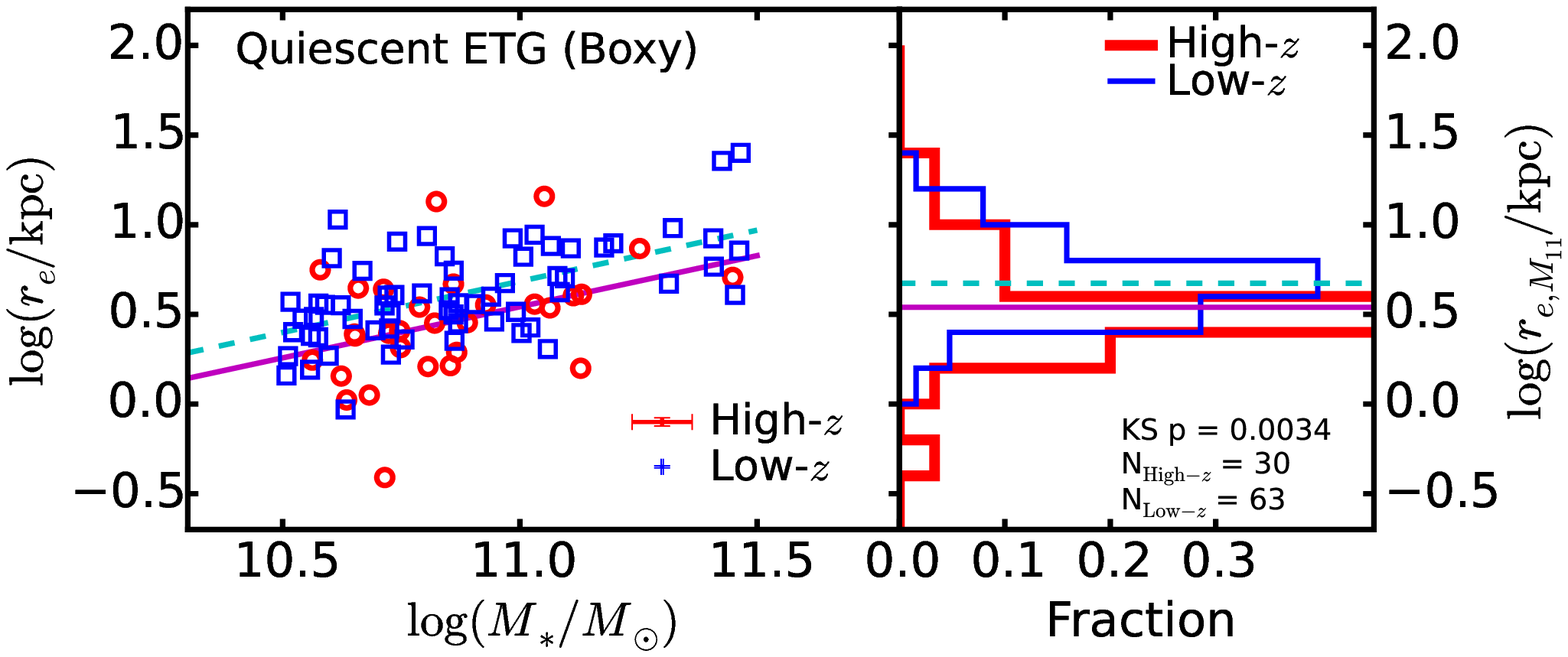}%
	\hfil 
	\includegraphics[width=8cm]{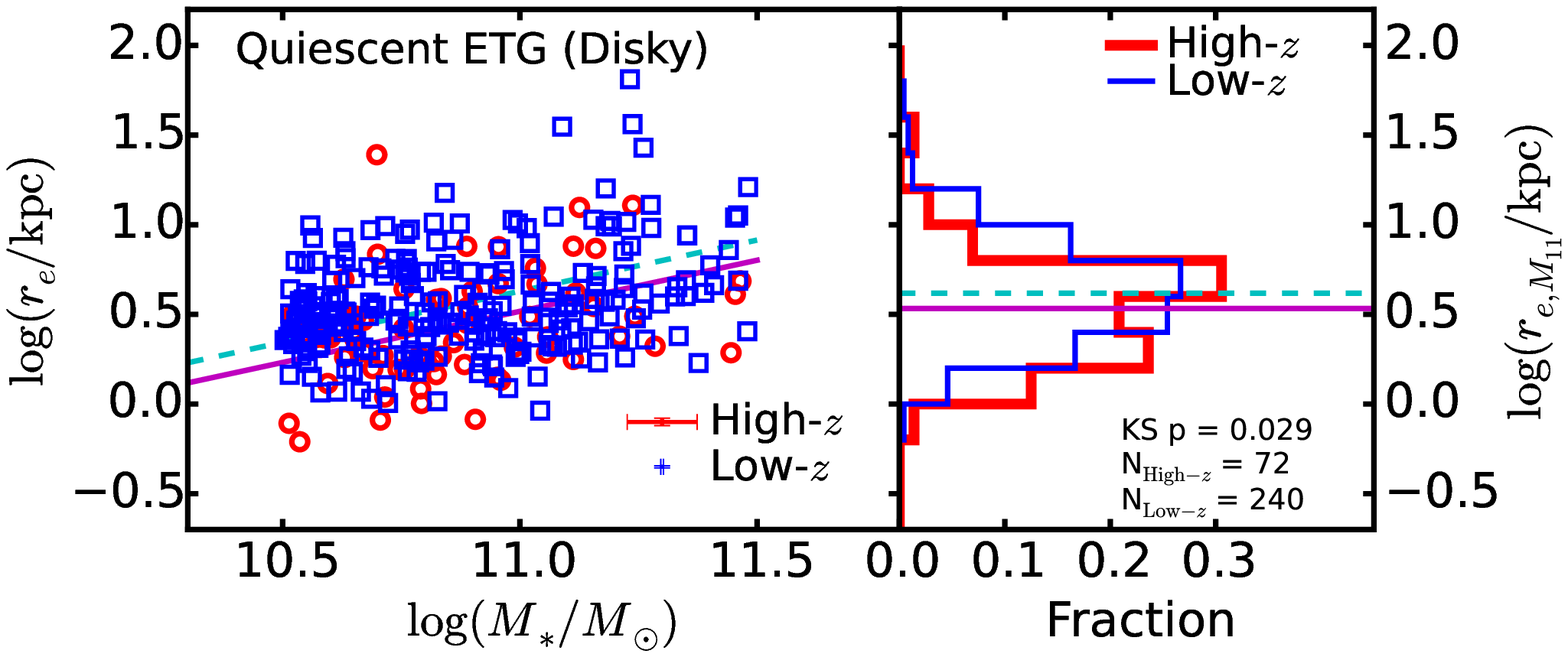}%
	\caption{
		Mass-size relation and histogram of the mass normalized size for boxy (left panels)
		and disky (right panels) ETGs plotted for less massive galaxies with $\log(M_{*}/M_{\odot}) < 11.5$.
		Symbols are the same as in Figure \ref{fig: M_Re_all}.
		\label{fig: M_Re_DB_LowMass}
	}
	\vspace{0\baselineskip}
\end{figure*}

As the disky fraction is consistent between $z\sim1$ and $0$,
we suggest that the disk-like ETGs found in massive clusters at $z\sim1$ \citep[e.g.,][]{Cerulo+14, De_Propris+15},
which should simply appear as disky ETGs in this study,
are transformed into disky ellipticals or S0s in the local universe rather than boxy ETGs.
To constrain the process responsible for the evolution, 
we compare the size mass relation between the high- and low-redshift samples separately for boxy and disky ETGs
within the stellar mass range of $10.5 < \log(M_{*}/M_{\odot}) \leq 11.5$.
In Figure \ref{fig: M_Re_DB_LowMass}, we plot the size mass relation and histograms of the mass normalized size $r_{e, M_{11}}$.
For boxy ETGs, the median sizes are 
$<\log(r_{e, M_{11}} / \rm{kpc})> = 0.67 \pm 0.03$ and $0.54 \pm 0.03$ for the low- and high-redshift samples, respectively.
For disky ETGs,
the median mass normalized sizes are
$<\log(r_{e, M_{11}} / \rm{kpc})> = 0.62 \pm 0.03$ and $0.53 \pm 0.06$ for the low- and high-redshift samples, respectively.
The KS test gives the p-value of $0.0034$ and $0.029$ respectively for boxy and disky which indicates the differences of the distributions between high- and low-redshift samples are significant.
%
We conclude that both the boxy and disky ETGs grow their size in $z<1$.
The process of the size growth for the ETGs should not be accompanied with the transformation of the isophote shapes and probably dynamical properties,
considering the constant disky fraction in $z<1$.
%

Considering the constant disky fraction,
we suggest that the main cause of the size growth and the morphological evolution of the intermediate mass ETGs in cluster environment may be less violent processes than mergers,
such as the accretion of low mass galaxies onto outskirts.
Even if mergers occur, the mass ratio should be enough large (i.e., the progenitor mass must be enough unequal) not to convert disky isophote shapes into boxy,
as major mergers \citep{Naab+99, Naab+03} or dry mergers with small mass ratios \citep{Khochfar+05, Naab+06} can convert disky ETGs into boxy.

We should be cautious about about the progenitor bias \citep[e.g.,][]{van_Dokkum+96, Saglia+10} i.e., blue star-forming disk galaxies at $z\sim1$ may enter into quiescent ETG sample at $z\sim0$.
If newly quenched star-forming disks with larger typical size appear as disky ETGs at $z\sim0$,
the disky fraction might be constant even if some part of disky ETGs at $z\sim1$ is converted into boxy,
and the size growth in disky ETGs between $z\sim1$ and $0$ may be observed.

To check whether newly quenched galaxy significantly contaminate the low redshift disky ETGs,
we have compared the color of the low-redshift ETGs between disky and boxy.
We compute $u-g$ color in two apertures, the Petrosian and the central 3-arcsec apertures.
If the low-redshift disky ETGs contain significant fraction of newly quenched galaxies,
the average color becomes bluer in disky than in boxy especially in the Petrosian apertures which contain lights from the outer disk with younger age than central bulge, assuming that the newly quenched disks become disky ETGs.
However, there is no significant difference in the average color and color distribution between disky and boxy ETGs in both Petrosian and 3-arcsec apertures.
Thus, we conclude that the low redshift disky ETGs are not contaminated by newly quenched disks,
and the evolution of the disky fraction is not affected by the progenitor bias.
This is also supported by the fact that the low redshift disky and boxy ETGs have similar sizes.


\section{Summary}
We measured the isophote shapes of ETGs in massive galaxy clusters at redshift $z\sim1$ and $0$
to investigate the evolution of the dynamical properties of ETGs.

We create high-redshift quiescent ETG sample residing in massive galaxy clusters at $z\sim1$
using imaging and spectroscopic data obtained in {\it HST} Cluster SN Survey \citep{Dawson+09}.
We selected spectroscopic members of the $z\sim1$ clusters.
Then, red galaxies are chosen based on the central $i_{775} - z_{850}$ color as quiescent galaxies.
Finally, ETGs are picked up from the quiescent galaxies using the concentration index and surface brightness.

We also prepare low-redshift quiescent ETG sample in massive clusters at $z\sim0$ using SDSS DR12 \citep{Alam+15} for comparison.
We selected quiescent galaxies using the color-magnitude diagram
with the help of the stellar mass-age and stellar mass-metallicity relation \citep{Thomas+05} and BC03 simple stellar population models \citep{Bruzual+03}.
Then, ETGs are selected based on the concentration index and surface brightness.

We develop an isophote shape analysis code which can be used for high-redshift galaxies with low surface brightness and small apparent sizes based on \citet{Bender+87}.
We have confirmed with a nearby galaxy that our code gives the consistent result to previous studies.
We also estimate the effect of PSF on the isophote shape measurement,
and find that the small axis ratio makes the measured $a_{4}$ value larger than intrinsic one
while the galaxy size and S\'{e}sic index only changes absolute value of the parameter.
As the PSF size in physical scales are similar between high- and low-redshift samples,
and as the axis ratios, galaxy sizes and S\'{e}rsic indices do not change very much between these redshifts,
we conclude that the effect of PSF does not introduce systematics to the evolution of the disky fraction.

Our results can be summarized as follows.
First, the parameter correlations between $a_{4}$ and other galaxy properties such as the stellar mass and size
are similar between $z \sim 1$ and $z \sim 0$ samples: ETGs tends to be boxy with increasing stellar mass or size.
Second, the main population of ETGs changes from disky to boxy at a critical mass of $\log(M_{*}/M_{\odot})\sim11.5$
regardless of the redshifts, with the massive end dominated by boxy galaxies.
Finally, we do not find significant difference in the disky to total ETG fraction between $z\sim1$ and $0$.

The fact that the massive ETGs ($\log(M_{*}/M_{\odot})\gtrsim11.5$) is basically boxy at $z\sim1$
indicates that the isophote shapes and probably the dispersion-dominated dynamics are characterized at higher redshifts.
The constant critical mass between $z\sim1$ and 0 is consistent with the mass quenching scenario
where shock-heated gas produced in a halo with $\log(M_{\rm{halo}} / M_{\odot}) \gtrsim 12$ and/or AGN feedback make mergers dry to produce boxy ETGs \citep[][and references therein]{Kormendy+09}.
This halo mass corresponds to the stellar mass of $\log(M_{\rm{*}} / M_{\odot})\sim11.2$ assuming baryon fraction of 1/6,
which is in good agreement with the critical mass of boxy-disky transition of the high- and low-redshift ETGs in our sample.

The environment may also play a role in characterizing isophote shapes and dynamical properties.
However, we need larger number of high-redshift ETG sample especially in outer region of galaxy clusters
to investigate redshift evolution in the disky fraction.

For less massive ETGs where disky galaxy is the dominant population,
the morphological evolution and size growth are reported in previous studies.
ETGs in massive clusters at $z\sim1$ tends to be more flattened, less concentrated, in other words, more disk-like, and small.
Considering the constant disky ETG fraction,
we suggest that disk-like ETGs, which should simply appear as disky ETGs in this study, in massive clusters at $z\sim1$
are transformed into disky ellipticals or S0s in the local universe.
The main cause of the size and morphological evolution
should be less violent processes than mergers
such as the accretion of low mass galaxies onto outskirts.
Even if mergers occur, the progenitor mass must be enough unequal not to convert disky isophote shapes into boxy,
as major mergers \citep{Naab+99, Naab+03} or dry mergers with small mass ratios \citep{Khochfar+05, Naab+06} can convert disky ETGs into boxy.

\acknowledgments
\paragraph{Acknowledgements}

We thank a very helpful referee for comments and suggestions that improved the quality of this paper.
We also thank Mitsuru Kokubo for number of discussions about the bias on the disky fraction.
KM thanks Masami Ouchi for helpful comments and advice.
KM aslo thanks Chris Lidman for encouraging comments.
%
Financial support for this work was provided by NASA through program GO-10496 from the Space Telescope Science Institute, which is operated by AURA, Inc., under NASA contract NAS 5-26555. This work was also supported in part by the Director, Office of Science, Office of High Energy and Nuclear Physics, of the U.S. Department of Energy under Contract No. AC02-05CH11231, as well as a JSPS core-to-core program “International Research Network for Dark Energy” and by a JSPS research grants (20040003, 23340041, 26287029).
This work was supported by JSPS Program for Advancing Strategic International Networks to Accelerate the Circulation of Talented Researchers.
The authors wish to recognize and acknowledge the very significant cultural role and reverence that the summit of Mauna Kea has always had within the indigenous Hawaiian community. We are most fortunate to have the opportunity to conduct observations from this mountain. Finally, this work would not have been possible without the dedicated efforts of the daytime and nighttime support staff at the Cerro Paranal Observatory.
Funding for SDSS-III has been provided by the Alfred P. Sloan Foundation, the Participating Institutions, the National Science Foundation, and the U.S. Department of Energy Office of Science. The SDSS-III web site is http://www.sdss3.org/.
SDSS-III is managed by the Astrophysical Research Consortium for the Participating Institutions of the SDSS-III Collaboration including the University of Arizona, the Brazilian Participation Group, Brookhaven National Laboratory, Carnegie Mellon University, University of Florida, the French Participation Group, the German Participation Group, Harvard University, the Instituto de Astrofisica de Canarias, the Michigan State/Notre Dame/JINA Participation Group, Johns Hopkins University, Lawrence Berkeley National Laboratory, Max Planck Institute for Astrophysics, Max Planck Institute for Extraterrestrial Physics, New Mexico State University, New York University, Ohio State University, Pennsylvania State University, University of Portsmouth, Princeton University, the Spanish Participation Group, University of Tokyo, University of Utah, Vanderbilt University, University of Virginia, University of Washington, and Yale University.
This research made use of NASA's Astrophysics Data System.
This research made use of the NASA/IPAC Extragalactic Database (NED) which is operated by the Jet Propulsion Laboratory, California Institute of Technology, under contract with the National Aeronautics and Space Administration.

{\it Facilities}:
{\it HST} (ACS),
Subaru (FOCAS),
KECK:II (DEIMOS),
VLT:Kueyen (FORS1),
VLT:Antu (FORS2),
Sloan (SDSS)

\bibliography{./qv}

\appendix

\section{A. Effects of the Difference of the Image Quality between Low- and High-Redshift Samples}
As image quality such as  the PSF size and signal-to-noise ratio ($S/N$) is different between high- and low-redshift samples,
it is important to assess how the difference of image quality affects our results.
What may matter is the difference of $S/N$ because the size of PSF is comparable in physical scale 
between low- and high-redshift samples (see Subsection\ref{Subsec: Loz_sample}).
In this section we present results of simulations to see how degrading of $S/N$ affects the measurement of some important parameters in this paper.
We degraded images of the low-redshift sample galaxies by adding Gaussian noise to make $S/N$ comparable to that of high-redshift ones,
and then measured morphological and structural parameters such as Gini coefficient, asymmetry, concentration index, surface brightness, effective radius, axis ratio, and S\'{e}rsic index
as well as isophote shape parameters such as $a_{4}$.

\subsection{A.1. Image Degradation of the Low-Redshift Galaxies}
We degrade images of the low-redshift sample galaxies in the following way.
First we determine the amount noise added onto the low-redshift galaxy images.
The surface brightness becomes fainter by $(1+z)^{-4}$ with increasing redshift due to cosmological dimming.
At the same time, galaxies become brighter as they have larger amount of bright young stars at higher redshift,
and the amount of luminosity evolution depends on star-formation and assembly history of galaxies.

As star-formation and assembly history is complicated, we need some assumptions on the luminosity evolution.
Here, we just adopt the difference of average luminosity between low- and high-redshift samples compared at the same stellar mass as the luminosity evolution.
In three mass ranges of $10.5\leq\log(M_{*})<11.0$, $11.0\leq\log(M_{*})<11.5$, $11.5\leq\log(M_{*})$,
we fit linear functions $M_{g\ \mathrm{or}\ z_{850}} = a (\log(M_{*}/M_{\odot}) - M_{0}) + b$, where $M_{0}$ is the lower limit for these mass ranges (i.e., $10.5, 11.0, 11.5$), $a$ and $b$ are the fitting parameters.
For low-redshift quiescent galaxies, we obtained the fitting functions, $M_{g} = -1.14 (\log(M_{*}/M_{\odot}) - M_{0}) -19.9$, $-1.43 (\log(M_{*}/M_{\odot}) - M_{0}) -20.5$, and $-1.66 (\log(M_{*}/M_{\odot}) - M_{0}) -21.2$ 
respectively for $10.5\leq\log(M_{*})<11.0$, $11.0\leq\log(M_{*})<11.5$, $11.5\leq\log(M_{*})$.
For high-redshift quiescent galaxies, we obtained the fitting functions, $M_{z_{850}} = -1.11 (\log(M_{*}/M_{\odot}) - M_{0}) -21.9$, $-0.896 (\log(M_{*}/M_{\odot}) - M_{0}) -22.7$, and $-1.31 (\log(M_{*}/M_{\odot}) - M_{0}) -23.3$.
As the difference of the parameter $b$ is $\sim2.0$ in all stellar mass bins, we adopt 2.0 mag as the luminosity evolution.
This is largely consistent with the luminosity evolution of $\sim2-3$ mag for galaxies with $\log(M_{*}/M_{\odot})=11-12$ predicted by BC03 SSP models assuming passive evolution with no merger.

Considering the luminosity evolution of 2.0 mag and cosmological surface brightness dimming between at $z\sim1.2$ ($SB_{\mathrm{dim}}=-2.5\log{(1+z)^{-4}}\sim3.4$) and $z\sim0.031$ ($SB_{\mathrm{dim}}\sim0.1$),
the surface brightness of the low-redshift galaxies in $g$ band should be observed brighter by 0.7 mag than that of the high-redshift galaxies in $z_{850}$.
For example, the surface brightness of 25.0 $\mathrm{mag\ arcsec^{-2}}$ in $z_{850}$ of a galaxy at $z\sim1.2$ is intrinsically 21.6 $\mathrm{mag\ arcsec^{-2}}$ without cosmological dimming.
As it would evolve to be fainter by 2.0 mag at low-redshift, the surface brightness would be 23.6 $\mathrm{mag\ arcsec^{-2}}$ in $g$ without cosmological dimming.
Finally, considering cosmological dimming at $z\sim0.031$, the observed surface brightness of the low-redshift galaxy is 23.7 $\mathrm{mag\ arcsec^{-2}}$ in $g$ band which is brighter by 1.3 $\mathrm{mag\ arcsec^{-2}}$ than the original surface brightness of 25.0 $\mathrm{mag\ arcsec^{-2}}$ in $z_{850}$.
We also make the pixel scale of the degraded images of the low-redshift galaxies (0.396 arcsec/pix $=0.24$ kpc/pix at $z=0.03$) comparable to that of the high-redshift galaxies (0.05 arcsec/pix $=0.41$ kpc/pix at $z=1.2$) by binning the images by $2\times2$ pixels.
Although the important spatial scale is the PSF size, this binning procedure makes the spatial sampling (including the PSF size as well as the pixel scale) comparable between low- and high-redshift samples.

We generate the {\it degraded} images of the low-redshift galaxies based on the discussion above.
The one sigma background noise level per pixel of our high-redshift sample is equivalent to $\sigma_{\mathrm{bkg, High-}z} = 24.7$ $\mathrm{mag\ arcsec^{-2}}$ in $z_{850}$ on average.
Therefore, if the background noise is $\sigma_{\mathrm{bkg, High-}z^{\prime}} = 23.4$ $\mathrm{mag\ arcsec^{-2}} = 2.22 \times 10^{-18}$ $\mathrm{erg s^{-1} cm^{-2} \AA^{-1} arcsec^{-2}} $,
the $S/N$ becomes comparable between high and low redshift at the surface brightness level.
Since $\sigma_{\mathrm{bkg, Low-}z}$ of the original $2\times2$-binned images of the low-redshift galaxies is
$1.05 \times 10^{-18}$ $\mathrm{erg s^{-1} cm^{-2} \AA^{-1} arcsec^{-2}} = 24.9$ $\mathrm{mag\ arcsec^{-2}}$,
we create the {\it degraded} images by adding Gaussian noise to the original images binned by $2\times2$ pixels.
One sigma of the added Gaussian noise corresponds to
$\Delta \sigma_{\mathrm{bkg}} = \sqrt{ \sigma_{\mathrm{bkg, High-}z^{\prime}}^{2} - \sigma_{\mathrm{bkg, Low-}z}^{2} } = 1.96 \times 10^{-18}$ $\mathrm{erg s^{-1} cm^{-2} \AA^{-1} arcsec^{-2}} $.

\subsection{A.2. Effects of the Image Degradation on the Morphological Parameters}
Using original and {\it degraded} images we measure the Gini coefficint \citep{Abraham+03}, asymmetry \citep{Abraham+96}, concentration index \citep{Doi+93}, and mean surface brightness \citep{Doi+93},
and investigate which parameters
are insusceptible to degrading of $S/N$ of an image.
We use interloper-subtracted images in this procedure.
We run \verb"SExtractor" and \verb"GALFIT" with the {\it degraded} images in the same manner as original to obtain the interloper-subtracted version of the {\it degraded} images.

We measure the Gini coefficients and asymmetry following \citet{Meyers+12}.
The Gini coefficient is measured within an quasi-Petrosian aperture.
First, preliminary aperture is prepared by collecting all pixels exceeding $1.0\sigma_{\mathrm{bkg}}$ above the background level
and contiguous to the center of the target, where $\sigma_{\mathrm{bkg}}$ is one sigma background noise per pixel.
We smooth the image with a Gaussian kernel of $\sigma=$2pixels when we take one-sigma isophote.
In the preliminary aperture, we compute the quasi-Petrosian flux \citep{Abraham+07} with the original image rather than smoothed one.
Then, the quasi-Petrosian aperture is created by collecting pixels exceeding the quasi-Petrosian flux in the preliminary aperture with the original image.
We calculate the Gini coefficient within this quasi-Petrosian aperture.
The error is estimated from bootstrap resampling of the pixels in the aperture \citep{Abraham+03}.
We randomly resample the same number of pixels from the aperture allowing overlaps, then recompute the Gini coefficient with the resampled pixels.
We recompute the Gini coefficient this way 1000 times to determine the probability distribution function of the Gini coefficient for each galaxy and record the standard deviation of this distribution as the Gini coefficient error.

The asymmetry with symmetrized aperture consisting of the intersection of the quasi-Petrosian aperture with its 180 deg rotation.
The rotation center is iteratively determined so that the measured asymmetry is minimized.
As background noise has some contribution to the measured asymmetry, we estimate the amount of contribution and subtract it from the measured asymmetry.
We generate 1000 artificial background images with Gaussian fluctuation with standard deviation of $\sigma_{\mathrm{bkg}}$ and measure their asymmetry in the same aperture as for the target.
We subtract the average of the 1000 asymmetry measurement of the background, and adopt the standard deviation as an estimate of error as the noise of our image is dominated by background.
How we measure the concentration index and mean surface brightness is described in Section \ref{Subsec: SelQETGs}.

\begin{figure}[htbp]
	\vspace{0\baselineskip}
	\begin{center}
	\includegraphics[width=17.0cm]{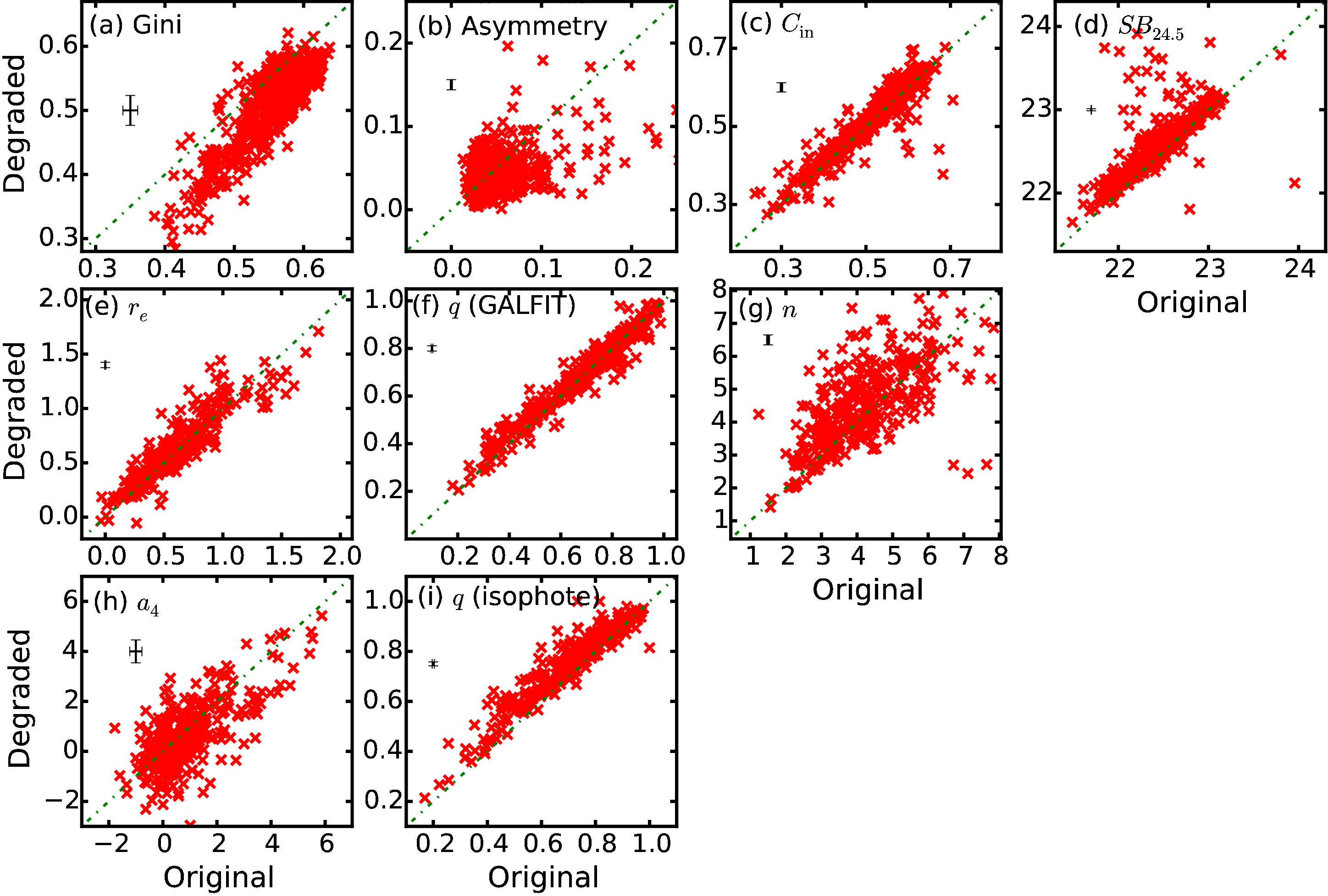}
	\caption{
		{\it Top panels}: 
		Comparison of the morphological parameters, the Gini coefficient (a), asymmetry (b), concentration index (c), and mean surface brightness (d),
		of the low-redshift quiescent galaxies between those measured in original image ($x$ axis) and in {\it degraded} images ($y$ axis).
		{\it Middle panels}:
		Comparison of the structural parameters, the effective radius (e), axis ratio (f), and S\'{e}rsic index (g) measured by GALFIT
		of the low-redshift quiescent ETGs between those measured in original image ($x$ axis) and in {\it degraded} images ($y$ axis).
		{\it Bottom panels}:
		Comparison of the isophote shape parameters, $a_{4}/a$ (h) and axis ratio (i) measured from isophote shapes
		of the low-redshift quiescent ETGs between those measured in original image ($x$ axis) and in {\it degraded} images ($y$ axis).
		For all panels, {\it green dash-dotted lines} represent $y=x$ for reference, and median measurement uncertainty is given by error bars.
		\label{fig:DGR_All}
	}
	\end{center}
	\vspace{0\baselineskip}
\end{figure}

In the top panels in Figure \ref{fig:DGR_All}, we compare the morphological parameters measured in original images and in {\it degraded} images.
The Gini coefficient tends to be systematically smaller and the asymmetry has large scatter when measured in the {\it degraded} images.
On the other hand, the concentration index $C_{in}$ is less affected by the image degradation in the systematic sense.
The mean surface brightness $SB_{24.5}$ is less affected than Gini coefficient and asymmetry although a few galaxies have larger value in the {\it degraded} images.
Considering this, we decided to make use of concentration and surface brightness for ETG selection in this paper (see Section \ref{Subsec: SelQETGs}).
The Gini coefficient is also useful for detecting multiple flux peaks \citep{Abraham+03}, e.g., merging systems,
but as we simply focus on ETGs, the flux concentration is enough for our purpose,
which is anther reason why we use the concentration rather than the Gini coefficient.
The surface brightness is a useful parameter to classify galaxy morphology as it appears in the Fundamental Plane \citep{Djorgovski+87}.
We note that as we have spectroscopic redshift, we can correct the cosmological surface brightness dimming safely.

\subsection{A.3. Effects of the Image Degradation on the Structural Parameters}
We investigate the effects of the image degradation on the structural parameters such as the effective radius, axis ratio, and S\'{e}rsic index measured by \verb"GALFIT"
presented in Subsection \ref{BP_Q_ETGs}.
In the middle panels in Figure \ref{fig:DGR_All}, we compare the structural parameters measured in original images and in {\it degraded} images.
While effective radii and axis ratios are not affected very much by the image degradation with small scatters,
S\'{e}rsic indices measured in {\it degraded} images have large scatters.

\subsubsection{A.4. Effects of the Image Degradation on the Isophote Shape Parameters}
We also investigate the effects on the image degradation on the isophote shape parameters presented in Section \ref{Sec: Results}.
In the bottom panels in Figure \ref{fig:DGR_All}, we compare the $a_{4}$ and axis ratios measured from isophote shape analysis.
The $a_{4}$ parameters have quite large scatter with small systematics $\Delta a_{4}/a \sim -0.2$\% compared to the uncertainty
when they are measured in the {\it degraded} images.
However, as large uncertainty affects the shape of the distribution of the $a_{4}$ parameter, the disky-to-total fraction is affected by
large measurement uncertainty, 
which is presented in the next subsection.
The axis ratio measured from isophote shape 
tend to be large by $\sim0.04$.


\section{B. The disky fraction affected by uncertainty of the \lowercase{$a_{4}$} parameter}
\begin{figure*}[htbp]
	\vspace{0\baselineskip}
	\begin{center}
	\hspace{-0.5cm}	
	\includegraphics[height=15.0cm]{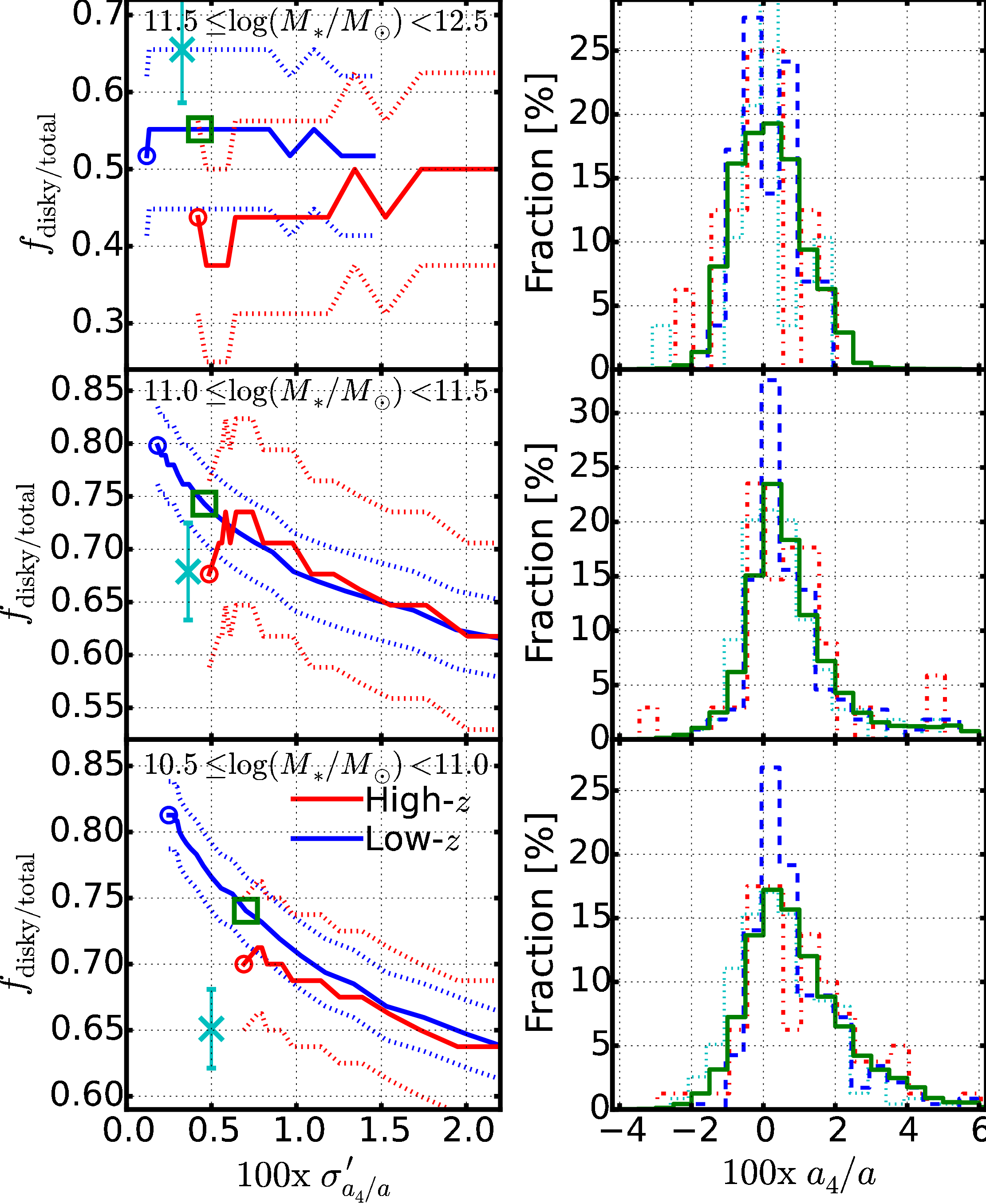}
	\caption{
		{\it Left}: 
		Simulated disky fraction of the high- ({\it red}) and low-redshift ({\it blue}) samples as functions of uncertainty of the $a_{4}$ parameter.
		We carried out 1000-time Monte Carlo resampling by adding Gaussian noise to obtain these plots.
		The measured disky fraction (without artificial noise added) is denoted by circles.
		{\it Green squares} represents the disky fraction of the low-redshift sample simulated with equivalent $a_{4}$ uncertainty to the high-redshift sample.
		{\it Cyan crosses} indicate the disky fraction of the {\it degraded} low-redshift sample.
		{\it Right}: 
		Histograms of the $a_{4}$ parameter of the high-  ({\it red dot-dashed lines}) and low-redshift ({\it blue dashed lines}) samples.
		{\it Green solid lines} represents the histograms of the low-redshift sample simulated with equivalent $a_{4}$ uncertainty to the high-redshift sample.
		{\it Cyan dotted lines} indicate the histograms of the {\it degraded} low-redshift sample.
		\label{fig:a4Frac_SIM}
	}
	\end{center}
	\vspace{0\baselineskip}
\end{figure*}

We present how the distribution of the $a_{4}$ parameter and the disky-to-total fraction are affected by the measurement uncertainty of $a_{4}$.
We have carried out Monte-Carlo simulations
in which the $a_{4}$ parameters of low- and high-redshift samples are resampled after Gaussian noises are added 
and then the disky fractions are computed in each resample.
We have investigated how the disky fraction changes as functions of the uncertainty of the $a_{4}$ parameter.
In the left panels of Figure \ref{fig:a4Frac_SIM},
we present the simulated disky fraction as functions of the uncertainty of the $a_{4}$ parameter.
Here the uncertainty (x-axis) is computed as the square root of the quadratic sum of the median of the original measurement uncertainty in a stellar mass bin
and sigma of the Gaussian noises added in the simulations.
For the lower stellar mass bins ($\log(M_{*}/M_{\odot}) < 11.5$), the disky fraction decreases with the increasing uncertainty. 
We also plot the disky fraction of the low-redshift sample measured with the {\it degraded} images with {\it cyan crosses} in the figure 
which are also smaller than the original value and comparable to the fraction 
of the high-redshift sample within uncertainty.

The decrease of the disky fraction is probably due to the fact that
the shape of the distribution of the $a_{4}$ parameter is modified by the uncertainty.
In the right panels in Figure \ref{fig:a4Frac_SIM}, 
we present the histograms of the $a_{4}$ parameter.
For the lower stellar mass bins ($\log(M_{*}/M_{\odot}) < 11.5$),
the histograms of the low-redshift sample simulated with equivalent $a_{4}$ uncertainty to the high-redshift sample ({\it green solid lines})
are larger than the original low-redshift sample ({\it blue dashed line}) in the boxy bins ($a_{4}/a \leq 0$).
This bias \citep[Eddinton bias, ][]{Eddington13, Teerikorpi04}
occurs when the distribution functions (here, the histograms of $a_{4}/a$) have second or higher order differential terms.
In the cases of right panels in Figure \ref{fig:a4Frac_SIM},
when the noises are added,
the number of galaxy moving from the disky side to the boxy side is larger than that from boxy to disky.
The disky fraction would asymptotically reach to 0.5 when we increase the added noises as the distribution would be flat.
Taking account of the Eddinton bias, i.e.,
if we compare the disky fractions of the low-redshift sample simulated with equivalent $a_{4}$ uncertainty to the high-redshift sample,
they become comparable to those of the high-redshift within uncertainty.
In order to detect possible differences of the disky fraction,
we have to increase the sample size of both the low- and high-redshift samples to $\sim$300, 400, and 800 for the highest, intermediate, and lowest mass bins.
In order to compare disky fraction without Eddington bias, we have to decrease the measured uncertainty of the $a_{4}$ parameter of the high-redshift sample, which will be possible with the next generation telescopes.

\end{document}